\documentclass[twocolumn,showpacs,preprintnumbers,amsmath,amssymb,prb]{revtex4-1}

\usepackage[active]{srcltx}
\usepackage{graphicx}% Include figure files
\usepackage{dcolumn}% Align table columns on decimal point
\usepackage{bm}% bold math

\renewcommand{\d}{{\bm{\delta}}}
\newcommand{\A}{{\bf A}}
\newcommand{\E}{{\bf E}}

\renewcommand{\j}{{\bf j}}
\renewcommand{\k}{{\bm{k}}}

\newcommand{\q}{{\bm{q}}}
\newcommand{\p}{{\bm{p}}}
\renewcommand{\r}{{\bm{r}}}

\renewcommand{\v}{{\bf v}}
\newcommand{\I}{{\rm i}}

\newcommand{\re}{\ensuremath{\mathrm{Re}\,}}
\newcommand{\im}{\ensuremath{\mathrm{Im}\,}}

\def\gsim{\lower.35em\hbox{$\stackrel{\textstyle>}{\textstyle\sim}$}}
\def\lsim{\lower.35em\hbox{$\stackrel{\textstyle<}{\textstyle\sim}$}}

\begin{document}

\title{Plasmonics in Dirac systems: from graphene to topological insulators}

\author{T. Stauber}

\affiliation{Departamento de F\'{\i}sica de la Materia Condensada, Instituto Nicol\'as Cabrera and Condensed Matter Physics Center (IFIMAC), Universidad Aut\'onoma de Madrid, E-28049 Madrid, Spain}

\begin{abstract}
The recent developments in the emerging field of plasmonics in graphene and other Dirac systems are reviewed and a comprehensive introduction to the standard models and techniques is given. In particular, we discuss intrinsic plasmon excitations of single and bilayer graphene via hydrodynamic equations and the random phase approximation, but also comment on double and multilayer structures. 
Additionally, we address Dirac systems in the retardation limit and also with large spin-orbit coupling including topological insulators. Finally, we summarize basic properties of the charge, current and photon  linear response functions in an appendix.
\end{abstract}
\pacs{78.67.Wj, 78.70.En, 42.25.Bs, 78.20.Ci}

\pacs{73.22.Pr,73.20.-r,79.20.Ws,79.60.Dp,78.47.J}

%\noindent {\it keywords}: optical properties of graphene, 2D plasmons

\maketitle

%%%%%%%%%%%%%%%%%%%%%%%%%%%%%%%%%%%%%%%%%%%%%%%%%%%%%%%%%%%%%
%  SECTION INTRODUCTION
%%%%%%%%%%%%%%%%%%%%%%%%%%%%%%%%%%%%%%%%%%%%%%%%%%%%%%%%%%%%%
 \section{Introduction}
The outstanding optical properties of two-dimensional (2D) carbon sheets were the key to the discovery of exfoliated graphene in 2004,\cite{Novoselov04,NovoselovPNAS05,Blake07,Neto09} and its optoelectronic properties are arguably the most promising ones for applications.\cite{Bonaccorso10,Avouris10,Koppens11} Especially, the large intrinsic carrier
mobilities and doping tunability have led to a number of proposals, where the
engineering of long-lived graphene plasmons could play a major
role.\cite{Rana08,Jablan09,Koppens11,Grigorenko12}

Plasmon excitations are intrinsic collective charge or current oscillations coupled via the Coulomb interaction which constitutes the restoring force. Obviously, the group velocity of these oscillations cannot exceed the velocity of light. This means that the plasmon dispersion lies outside the light cone in the near-field (evanescent) regime. The group velocity only merges with the velocity of light for low energies $\hbar\omega\lesssim\alpha E_F$ due to retardation effects where 
\begin{equation}
\alpha=\frac{e^2}{4\pi\varepsilon_0\hbar c}\approx\frac{1}{137}
\end{equation}
denotes the fine-structure constant with $\varepsilon_0$ the vacuum permittivity and $E_F=\hbar v_Fk_F$ the Fermi energy of the doped graphene layer. At THz frequencies, the confined collective oscillations of electrons thus enable the manipulation of electromagnetic energy at sub-wavelength scales which is usually coined as plasmonics.

The field of plasmonics in nobel metals has already attracted a great deal of attention for the past 15 years,\cite{Ebbesen98,GarciaVidal10,Brongersma10,Stockman11,Maier07,Klimov12} where the  collective charge excitations on the metallic surface localize the electromagnetic field on sub-wavelength dimensions and can act as strong dipole or antenna. Additionally, the field enhancement can become very large such that single molecules are being detected by Raman scattering.\cite{Maier06} This opened up the possibility to efficiently couple light to electrons and thus merging photonics and electronics at nanoscale dimensions,\cite{Ozbay06} and already gave rise to a number of metamaterials.\cite{Veselago68,Pendry00,Engheta06,Zheludev12}

Plasmons in graphene provide a suitable alternative to noble-metal surface plasmon polaritons because of atomistic confinement of the electrons and the accompanied electromagnetic fields, relative long propagation lengths compared to the plasmon wavelengths and, most importantly, tunability.\cite{Rana08,Jablan09,Koppens11} Here, we will review the recent advances in the emerging field of graphene plasmonics and give a comprehensive introduction to the basic theoretical models and techniques.
					
\begin{figure}[t]
\begin{center}
  \includegraphics[angle=0,width=\columnwidth]{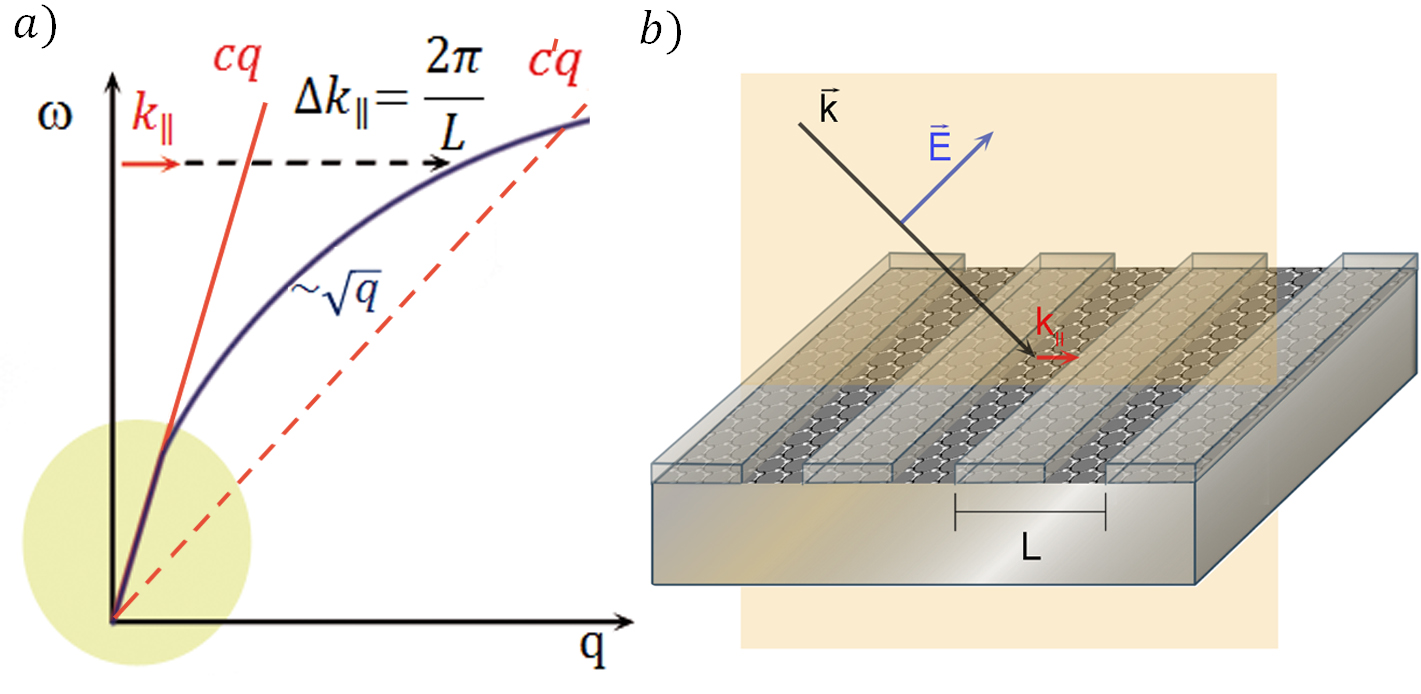}
\caption{(color online): a) Schematic 2D plasmon dispersion (blue curve) together with the light cone in vacuum (red solid line) and in an optically denser medium (red dashed line). The retardation regime is indicated by the circled region where strong light-matter interaction sets in. b) Excitation of plasmonic modes by light is possible via an artificial sub-wavelength periodicity providing the missing momentum, $\Delta k_\parallel$, of the incident electromagnetic radiation with parallel momentum $k_\parallel$.} 
  \label{figure1}
\end{center}
\end{figure}

Plasmons cannot directly couple to propagating electromagnetic radiation because the conservation of momentum is not satisfied in the photon absorption process. To couple light to surface plasmon polaritons on nobel metals, the Otto\cite{Otto68} or Kretschmann\cite{Kretschmann71} configuration is used, i.e., the velocity of light of the incoming light $c'$ is reduced by a factor of 2-10 due to an optically dense medium. The slope of the light cone is thus smaller ($c'<c$) and plasmons can be excited by incoming light under an appropriate incident angle, see Fig.  \ref{figure1}a).

For graphene on a typical dielectric substrate with relative dielectric constant $\epsilon\approx3$ and covered by air, the plasmon dispersion is shifted to larger $q$-vectors, approximated by the compact formula
\begin{equation}
\label{PlasmonIntro}
\frac{q}{q_{p}}\approx\alpha\frac{E_F}{\hbar\omega}
 \end{equation}
where $\omega=cq$ is the energy dispersion of the vacuum light cone, $q_{p}$ is the wave number of the plasmon and $E_F$ the Fermi energy of the doped graphene layer. For typical Fermi energies of $E_F=0.3$eV, there is thus a strong reduction of the wavelength even in the THz-regime ($\hbar\omega\gtrsim 4$meV), and the Otto or Kretschmann configuration can usually not be used to detect graphene's intrinsic plasmonic excitations.  Graphene plasmons were, therefore, first investigated by means of electron energy-loss spectroscopy (EELS) where the electronic beam was carrying the necessary momentum.\cite{Liu08,Eberlein08}

One possibility to couple propagating light to graphene's density oscillations is to break the translational symmetry. The necessary (missing) momentum is then provided by a patterned 2D surface with a periodic sub-wavelength structure, see Fig. \ref{figure1}b). For graphene, this was first achieved with a grated geometry revealing plasmon resonances with remarkably large oscillator strengths even at room temperature.\cite{Ju11} To this end, the absorption spectrum was obtained for two different polarizations: for the electric field parallel to the grating, the usual Drude peak was seen whereas a (plasmon) resonance showed up at finite $q$ in the case of perpendicular polarization.\cite{Ju11,Yan13} Alternatively, a graphene disk array can be used to excite plasmon resonances by incoming light.\cite{Yan12}

Graphene plasmons can also be excited without  periodic sub-wavelength patterning by converting the far-field modes into near-field modes via dipole scattering. By this, graphene plasmons were recently launched and detected directly by evanescent waves produced by illuminating an atomic force microscope (AFM) tip with propagating light in the infrared regime.\cite{Chen12,Fei12,Fei11} By this technique, also the propagation of plasmons can be studied and a new near-field scattering microscopy is developing which is likely to be superior to AFM scanning techniques for large sample areas. 

Up to now, the experimental techniques can only access the low wave number regime $q\sim10^4-10^5$cm${}^{-1}$ for which the standard theoretical tools such as hydrodynamic models or the random phase approximation (RPA) are well justified (assuming usual doping levels with $q/k_F\ll1$). In this topical review, we want to summarize these basic theoretical concepts needed to describe the light-graphene interaction for various physical environments and conditions. We further intend to illustrate the same ideas from different perspectives and will thus derive the basic formulas using various approaches.

In Sec. II, we present the fundamental formulas concerning the 2D plasmon dispersion using phenomenological descriptions. In Sec. III and IV, we review various aspects of intrinsic plasmon excitations of single layer and bilayer graphene, mostly based on the RPA. In Sec. V, we derive the  plasmonic spectrum of electrostatically coupled graphene layers and in Sec. VI, we discuss the effect of retardation on longitudinal (or transverse magnetic - TM) as well as on  transverse (or transverse electric - TE) plasmons. We close with an account on plasmon excitations in Dirac systems with large spin-orbit coupling. In an appendix, we summarize basic concepts and results of linear response theory for Dirac Fermions based on the hexagonal tight-binding model as well as for the electromagnetic gauge field.

\section{Hydrodynamic models}
\label{HydrodynamicModels}
Plasmons are collective density oscillations present in almost all electronic systems. They are straightforwardly obtained via a hydrodynamic description based on the continuity equation and linear response
\begin{equation}
\label{HD}
-\I\omega\rho=-\nabla\cdot\j=\chi_{jj}^+\nabla\cdot\A=-\frac{\chi_{jj}^+}{\I\omega}\nabla^2\phi\;,
\end{equation}
where we introduced the (local) longitudinal current response function, $\chi_{jj}^+=\chi_{jj}^+(\omega)$, defined in the appendix, Eq. (\ref{CurrentCurrent}), and used $\E=\I\omega\A=-\nabla\phi$ with $\E$ the electric field and $\A,\phi$ the vector and electrostatic potential, respectively. With the Fourier transform, $\phi(\r)=\phi_\q e^{\I \q\cdot\r}$, and the general relation between the potential and the charge density $\phi_\q=v_q\rho_\q$, we can write the above equation as
\begin{equation}
(\omega^2-\chi_{jj}^+v_q q^2)\rho_\q=0\;.
\end{equation}
Collective density oscillations are thus defined by the dispersion relation $\omega_p^2=\chi_{jj}^+v_qq^2$ which holds for all dimensions. This approach is well justified in the long-wavelength limit $q\to0$ where the system can be described by an electron liquid. Including additionally pressure and shear forces will lead to the same plasmonic dispersion as obtained from the random phase approximation (RPA), discussed in the next section.\cite{Giuliani05}

We will now limit the following discussion to two dimensions (2D) and write the dispersion relation in terms of the Drude weight defined by $D=e^2\chi_{jj}^+(\omega\to0)$. This yields the usual expression $D_{2DEG}=e^2n/m$ for a 2D electron gas (2DEG) with particle density $n$ and electron mass $m$. Note that we have taken the local approximation $q\to0$ before the static limit $\omega\to0$ since the electrons cannot establish local equilibrium and remain dynamical. The other order of limits would be related to the density of states of the electron system and thus an equilibrium property.

With the 2D Coulomb interaction $v_q=\frac{e^2}{2\varepsilon_0\epsilon q}$ and $\epsilon$ the relative (effective) dielectric constant, the plasmon dispersion for a general 2D system in the local approximation is thus given by
\begin{equation}
\label{TwoDPlasmons}
\omega_p=\sqrt{\frac{D}{2\varepsilon_0\epsilon}q}\;,
\end{equation}
yielding the characteristic square-root dispersion in 2D. For graphene on the interface of two different dielectric media, one further has $\epsilon=(\epsilon_1+\epsilon_2)/2$, see Eq. (\ref{PlasmonDisRet}).

In the local limit, $q\rightarrow0$, we can treat general isotropic systems with energy dispersion relation $E(\k)\sim|\k|^\nu$ on the same footing, see appendix. For low temperature, one obtains the general result in terms of the chemical potential $\mu$,
\begin{equation}
\label{localCurrentResponse}
\chi_{jj}^+=\frac{g_sg_v\nu}{2}\frac{\mu}{2\pi\hbar^2}\;,
\end{equation}
with $g_s$, $g_v$, the spin- and  valley degeneracies, respectively. 

The 2D particle density is independent of the energy dispersion and given by $n=\frac{g_sg_v}{4\pi}k_F^2$ with $k_F$ the Fermi wave number. This results in a different density behavior of $\chi_{jj}^+$ for different $E(\k)\sim|\k|^\nu$. Assuming further a dependence of the dispersion on the momentum $p=\hbar k$, it is clear that the Drude weight will depend on $\hbar$ for all $\nu\neq2$.\cite{DasSarma09} Interestingly, for the two most prominent cases, monolayer graphene and a 2DEG, we have $\chi_{jj}^+=\frac{\mu}{\pi\hbar^2}$ and thus the same plasmon dispersion in terms of the chemical potential; only the density behavior is different.

For Dirac Fermions with $\nu=1$, the Drude weight can be written as
\begin{equation}
D_{Dirac}=\frac{e^ 2v_F}{\hbar}\sqrt{\frac{g_sg_vn}{4\pi}}=\frac{4\mu}{\pi\hbar}\sigma_0\;,
\end{equation}
with $\sigma_0=\frac{g_sg_v}{16}\frac{e^2}{\hbar}$ the universal conductivity.
The general plasmon dispersion of Eq. (\ref{TwoDPlasmons}) can further be expressed with respect to dimensionless quantities as  
\begin{equation}
\label{DiracPlasmons}
\frac{\hbar\omega_p}{\mu}=\sqrt{\frac{g_sg_v\alpha_g}{2\epsilon}}\sqrt{\frac{q}{k_F}}\;,
\end{equation}
with graphene's fine-structure constant $\alpha_g=\alpha\frac{c}{v_F}\approx2.2$. For $g_s=g_v=2$ and $\epsilon=2$, one obtains the formula of the introduction, Eq. (\ref{PlasmonIntro}).

\subsection{Drude model}
Dissipative effects on the plasmon dispersion are most easily included within the phenomenological Drude model which provides the corresponding conductivity. We will first recapitulate the results for 2D surface plasmon polaritons emerging on the surface of nobel metals. We then discuss genuine 2D plasmons, i.e., graphene plasmons.

\subsubsection{Surface plasmon polaritons}
On metal/insulator interfaces, i.e., on a metal surface with negative dielectric constant covered by a dielectric medium with $\epsilon>0$, surface plasmon polaritons can exist up to the frequency $\omega_{spp}=\omega_p^{3D}/\sqrt{1+\epsilon}$, with the volume plasma frequency $\omega_p^{3D}=\sqrt{ne^2/\varepsilon_0m}$.\cite{Novotny06} Dissipation is introduced via the three-dimensional (3D) Drude model which leads to the following local dielectric function:\cite{Ashcroft76}
\begin{equation}
\epsilon_{3D}(\omega)=1-\frac{(\omega_p^{3D})^2}{\omega(\omega+\I\gamma)}\;,
\end{equation}
where $\gamma$ denotes the electronic relaxation rate. For silver, one finds  $\hbar\omega_p^{3D}=9.176$eV and $\hbar\gamma=21$meV and $\hbar\omega_p^{3D}=9.062$eV and $\hbar\gamma=70$meV for gold.\cite{Johnson72} For energies in the visible regime, also the response of bound electrons and high-energy interband transitions needs to be taken into account.\cite{Novotny06} Nevertheless, the plasmon dispersion is fixed by the bulk properties of the underlying 3D metal which cannot easily be changed. This is one of the main disadvantages compared to graphene's genuine 2D plasmons.
\subsubsection{Graphene plasmons}
Graphene's plasmons are intrinsic excitations of a truly 2D system. They are defined by the following "local" 2D dielectric function obtained from the Maxwell equations:
\begin{equation}
\label{TwoDDielectric}
\epsilon_{2D}(q,\omega)=\epsilon+\frac{\I\sigma(\omega) q}{2\omega\varepsilon_0}\;,
\end{equation}
with the 2D (local) Drude conductivity given by $\sigma(\omega)=\I e^2\chi_{jj}^+(\omega)/(\omega+\I\gamma)$. Using Eq. (\ref{TwoDPlasmons}), we thus have a similar expression for the dielectric function as in the 3D case,
\begin{equation}
\epsilon_{2D}(q,\omega)=\epsilon\left(1-\frac{\omega_p^2}{\omega(\omega+\I\gamma)}\right)\;.
\end{equation}
Genuine 2D plasmons are defined by $\epsilon_{2D}(q,\omega_p)=0$ and with the electronic relaxation time $\tau=1/\gamma$ this yields
\begin{equation}
\label{PlasmonPheno}
\omega_p^\tau=-\I\frac{1}{2\tau}+\sqrt{\omega_p^2-\frac{1}{(2\tau)^2}}\;.
\end{equation}

Given the relaxation time due to Coulomb\cite{Adam07} or resonant\cite{StauberMidGap07} scattering, the damping rate of graphene plasmons can be estimated. For graphene on a substrate, one usually sets $\hbar\gamma=10$meV. For suspended graphene, the main scattering mechanism at finite temperature is given by flexural phonons which can be eliminated by applying strain or placing graphene on a substrate.\cite{Castro10} We finally note that transport lifetimes are usually calculated in the local limit, but for high-frequency plasmons the $q$-dependence of the response function can become important.\cite{Principi13b}

\subsection{Semiclassical Boltzmann equation}
\label{sec:Euler}
The above treatment holds for any 2D electronic system and graphene's characteristic properties only entered through the local response function, i.e., $\omega_p$ and the phenomenological relaxation time $\tau$. We will now discuss hydrodynamic (Euler) equations which explicitly take into account the linear Dirac dispersion. We can also allow for electron as well as for hole currents and include damping terms defined via microscopic collision integrals.  

To derive the Euler equations for graphene, we follow Ref. \onlinecite{Ryzhii09}. For a general discussion including also a magnetic field and the full band structure, see Refs. \onlinecite{Roldan13b,Hanson13}. Starting point is the continuity equation for the semiclassical distribution function $\frac{d}{dt}f_{\k(t)}(\r,t)+\nabla\cdot\j_\k(\r,t)=0$ with the current given by $\j_\k=\v_\k f_\k$. For graphene, the velocity reads $\v_\k=v_F\k/k$ and with the equation of motion $\dot{\p}=e\nabla\phi$ we arrive at the collision-free Boltzmann equation for graphene,
\begin{equation}
\frac{\partial f}{\partial t}+v_F \frac{\p}{p}\frac{\partial f}{\partial \r}+e\frac{\partial \phi}{\partial \r}\frac{\partial f}{\partial \p}=0\;.
\end{equation}
 Expanding the Fermi distribution, $f\approx f^0-\frac{\partial f^0}{\partial \epsilon}(\p\cdot\v)$, multiplying the above equation with $\p$ and integrating over the phase space $d\Gamma_\p=\frac{g_sg_vd^2p}{(2\pi\hbar)^2}$, we arrive at the Euler equations for graphene as discussed by Ryzhii and co-workers:\cite{Ryzhii09}
\begin{equation}
\frac{3}{2v_F}\frac{\partial \langle p\rangle\v}{\partial t}+\frac{v_F}{2}\frac{\partial \langle p\rangle}{\partial \r}-ne\frac{\partial \phi}{\partial \r}=0
\end{equation}
with $\langle p\rangle=\int d\Gamma_\p p f_0$. The Euler equation has to be solved together with the continuity equation $\frac{\partial}{\partial t}n+\frac{\partial }{\partial \r}(n\v)=0$. 

The above equations hold for electrons as well as for holes and the effect of disorder, phonons and/or Coulomb interaction can be included by appropriate collision integrals. By linearization, one obtains analytical solutions for two limits: the symmetric bipolar and the monopolar system. In both cases, the plasmon dispersion has the same analytic structure as in Eq. (\ref{PlasmonPheno}), but the relaxation time is now replaced by expressions involving the collision integrals. In the monopolar case,  this leads to a square-root or linear plasmon dispersion, depending  on the screening behavior (of the gate), see Sec. \ref{Sec:AcousticPlasmons}. The symmetric bipolar system can support plasmons with a sound velocity $v_s\approx0.6v_F$, emerging due to the co-directional motion of electrons and holes. A novel plasmonic mode in neutral graphene due to excitontic effects was also reported in Ref. \onlinecite{Gangadharaiah08}, with sound-velocity $v_s=(1-e^{-N})v_F$ and $N=4$ the number of fermion flavors.

Within the above Euler equations, one can also discuss the generation of plasma waves by a dc current,\cite{Tomadin13} which was first proposed by Dyakonov and Shur in a 2DEG,\cite{Dyakonov93} and recently investigated experimentally in the context of graphene which opens up the possibility of effective THz generation.\cite{Vicarelli12} 

\section{Plasmons in single-layer graphene}
A plasmon is an oscillating charge density mode which is necessarily accompanied by a corresponding electric potential, neglecting retardation effects for the moment. Density and electric potential are related via the Poisson equation and the oscillations are thus sustained by the Coulomb interaction between electrons.

In order to describe plasmon excitations, the response of an electronic system to the total (screened) electric potential $\phi_{total}(\bm{r},t)$ is needed, which shall be denoted by $\chi_{\rho\rho}$. But it is often more convenient to discuss the total response of the system ($\chi_{\rho\rho}^{total}$) to the external potential, $\phi_{ext}(\bm{r},t)$. The long-ranged Coulomb interaction then needs to be treated self-consistently, leading to the following total density response: 
\begin{equation}
\label{RPAdensity}
\chi_{\rho\rho}^{total}(\q,\omega)=\frac{\chi_{\rho\rho}(\q,\omega)}{\epsilon(\q,\omega)}=\frac{\chi_{\rho\rho}(\q,\omega)}{1-v_q\chi_{\rho\rho}(\q,\omega)}\;,
\end{equation}
with the 2D Fourier transform of the electron-electron interaction $v_q=\frac{e^2}{2\varepsilon_0\epsilon q}$. The plasmonic excitations are then defined by the zeros of the dielectric function $\epsilon(\q,\omega)$, that also relates the total (screened) electric potential  to the externally applied potential via $\phi_{total}(\bm{q},\omega)=\phi_{ext}(\bm{q},\omega)/\epsilon(\bm{q},\omega)$. A plasmon is, therefore, a finite solution $\phi_{total}(\bm{q},\omega)$ that requires no external driving, i.e., it is self-sustained.

So far, the above analysis is exact. Taking now the response function $\chi_{\rho\rho}$ as the bare response without including vertex corrections, this is usually coined as the random phase approximation (RPA), $\chi_{\rho\rho}^{total}\to\chi_{\rho\rho}^{RPA}$, and represents the standard approximation to analyze the plasmonic spectrum within linear response theory. It is well justified in the high-density limit or for wave numbers $q\lesssim k_F$, with $k_F$ the Fermi wave number.\cite{Fetter03,Mahan00,Giuliani05}

Including retardation effects, the full current response needs to be considered. Assuming an isotropic system, the longitudinal (+) and transverse (-) current generated by a (total) gauge potential in linear response is given by $j^\pm=-q_e\chi_{jj}^\pm A_{total}^\pm$ with $q_e=-e$ the electron charge. The total gauge potential consists of the external potential and the field produced by the generated current, $A_{total}^\pm=A_{ext}^\pm+\Delta A^\pm$ and in linear response, we have $\Delta A^\pm=-q_ed^\pm j^\pm$ with $d^\pm$ the 2D photonic propagator, see appendix. The total response defined through $j^\pm=-q_e\chi_{jj,total}^\pm A_{ext}^\pm$ is thus given by
\begin{equation}
\label{chiTotal}
\chi_{jj,total}^\pm=\frac{\chi_{jj}^{\pm}}{1-q_e^2d^\pm\chi_{jj}^\pm}\;. 
\end{equation}
Again, we have $\chi_{jj,total}^\pm\to\chi_{jj,RPA}^\pm$ in the case of a vertex-free (bare) current response.

In the following, we will summarize the basic results for the plasmonic excitations based on the RPA in single layer graphene and also comment on various extensions. The underlying response functions are discussed in the appendix.

\subsection{Gappless Dirac Fermions}
Graphene is a 2D crystal where the carbon atoms form a hexagonal lattice. The two equivalent atoms in the unit cell give rise to two electronic bands which touch each other at the corners of the Brillouin zone. They can be grouped together to two inequivalent $K$-points which are related via time-reversal symmetry and around these $K$-(Dirac) points, the energy dispersion is conical and isotropic with Fermi velocity $v_F\approx c/300$.\cite{Neto09} Within this Dirac cone approximation, the response functions are also isotropic and we may drop the vector character of $\q\to q$.

The general, non-retarded plasmon dispersion including phenomenological damping is then defined by $\epsilon(q,\omega
_{p}-i\gamma )=0$, where $\gamma $ is the decay rate of the plasmons.\cite{Fetter03} For
weak damping, the plasmon dispersion $\omega_{p}(q)$ and the decay rate $%
\gamma $ are determined by
\begin{equation}
1=v_{q}\re\chi_{\rho\rho}(q,\omega _{p})\,, \quad\gamma
 =\frac{\im\chi_{\rho\rho}(q,\omega _{p})}{\frac{\partial }{\partial
\omega }\re\chi_{\rho\rho}(q,\omega )\left. {}\right\vert _{\omega
_{p}}}\;. \label{eq:gamma}
\end{equation}%

Solutions to the first equation require $\re\chi_{\rho\rho}>0$, which for single layer graphene is only the case for finite doping $E_F>0$ and $\omega >v_Fq$. Furthermore, a stable solution demands Im$\chi_{\rho\rho}=0$, which is the regime indicated by the white triangle of Fig.~ \ref{fig:Diagram}. In RPA, $\epsilon\to\epsilon_{RPA}$, this yields stable $\delta$-like excitations with an energy dispersion given in Eq. (\ref{DiracPlasmons}) for $q\lesssim k_F$.\cite{Wunsch06,Hwang07} For larger $q\gtrsim k_F$, the plasmon dispersion enters the regime of interband transitions (violet region of Fig.~ \ref{fig:Diagram}) where the plasmon becomes (Landau) damped due to dissipation into particle-hole excitations. This leads to a nonzero decay rate $\gamma$.\cite{Wunsch06} 

\begin{figure}
   \centering
   \includegraphics[height=0.465\columnwidth]{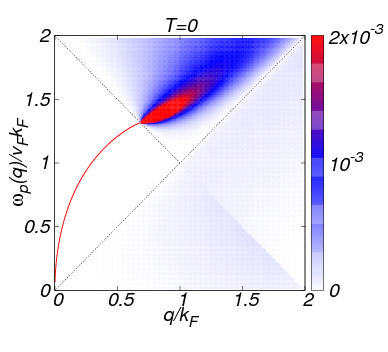} 
   \includegraphics[height=0.465\columnwidth]{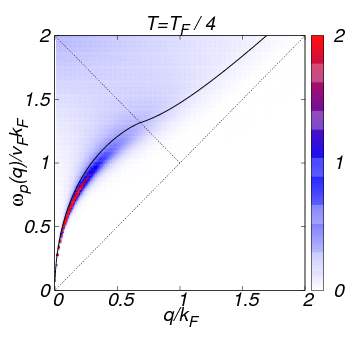} 
      \caption{Generalized loss function $S(q,\omega)=-\im \chi_{\rho\rho}^{RPA}(q,\omega+\I0)$ for doped graphene in units of $E_F/\hbar^2$ at zero temperature (left) and at $T=T_F/4$ (right). The region of undamped plasmons at $T=0$ is defined by straight lines. The black curve on the right hand side corresponds to $T=0$.}
   \label{fig:plasmonsSingle}
\end{figure}

In Fig. \ref{fig:plasmonsSingle}, we plot the generalized loss function $S(\bm{q},\omega)=-\im \chi_{\rho\rho}^{RPA}(\bm{q},\omega)$ indicating intrinsic plasmon excitations, see Sec. \ref{EnergyLoss}. The left hand side shows the dispersion at zero temerpature and the full red line stands for $\delta$-like undamped excitations which merge into the Landau-damped regime of interband transitions. 

\subsubsection{Finite temperature}
At finite temperature, plasmons can be sustained even by undoped graphene,\cite{Vafek06} i.e., the thermally activated charge density leads to coherent oscillations which are only weakly damped by the temperature induced interband transitions. There is a simple analytic expression for the energy dispersion using the formula of $\chi_{\rho\rho}$ for finite Fermi energy $E_F$ at $T=0$ by replacing $E_F\rightarrow2\ln2k_BT$.\cite{Vafek06,Falkovsky07} This substitution also holds for bilayer graphene.\cite{Stauber12}

For finite chemical potential, no closed analytic formula for $\chi_{\rho\rho}$ is known. Still, there is a compact expression involving only a one-dimensional integral, first obtained for the density response.\cite{Ramezanali09} The full current-current correlation was also derived, displaying a similar symmetry between the longitudinal and transverse channel as for the $T=0$ result, see Eq. (\ref{CurrentCurrentMu}).\cite{Gutierrez13}

On the right hand side of Fig. \ref{fig:plasmonsSingle}, the energy loss function displaying the plasmonic resonances is shown at finite temperature $T=T_F/4$ with the Fermi temperature $T_F=E_F/k_B$. The black line indicates the plasmon dispersion at $T=0$ obtained by Eq. (\ref{eq:gamma}) with $\gamma=0$. The plasmonic resonances are red shifted with respect to the $T=0$ result, but for larger temperature $T\gtrsim T_F/2$, they become blue shifted.

\subsubsection{Local response}
For small wave numbers $q\ll k_F$, the local response is sufficient for the description of the plasmonic excitations which is the case in most experimental setups. The current response function is thus often approximated by the constant Drude weight $D=e^2\chi_{jj}^+(\omega\to0)$ or, in terms of the conductivity, by $D=e^2\lim_{\omega\to0}\omega\im\sigma(\omega)$. This yields the expressions obtained from hydrodynamic models, see Eq. (\ref{TwoDPlasmons}).

The local approximation can be improved by also including the frequency dependence of the local conductivity. One can then split up the contribution in intra- and interband processes
\begin{equation}
\sigma(\omega)=\sigma_{intra}(\omega)+\sigma_{inter}(\omega)\;.
\end{equation}
Intraband processes lead to longitudinal, interband processes to transverse plasmons. 
The local conductivity has been discussed by numerous authors including magnetic fields, phenomenological disorder and finite temperature.\cite{Ando02,Peres06,Gusynin06,Gusynin07,Falkovsky07,Peres08}

\subsubsection{Undoped graphene}
The charge response function of undoped graphene was already calculated in 1994,\cite{Gonzalez94} and yields the characteristic square root singularity at  the one-particle energy dispersion $\omega=v_Fq$, discussed in the appendix. The conductivity is then given by 
\begin{equation}
\label{ConductivityNeutral}
\sigma^{\mu=0,T=0}(\omega,q)=\sigma_0\frac{\omega}{\sqrt{\omega^2-(v_Fq)^2}}\;,
\end{equation}
with $\sigma_0=\frac{g_sg_ve^2}{16\hbar}$ the universal conductivity. The conductivity $\sigma$ is real for $\omega>v_Fq$ and there are no plasmon excitations at zero temperature. But including vertex corrections in the polarizability leads to a positive imaginary part and a linear plasmon mode with sound velocity below the Fermi velocity emerges.\cite{Gangadharaiah08} Undoped graphene can also sustain plasmonic oscillations when exposed to circularly polarized external electric fields.\cite{Busl12}

\subsubsection{Beyond the Dirac cone approximation}
Up to now, the (bare) density response $\chi_{\rho\rho}$ was calculated within the Dirac cone approximation.
But for large Fermi energies with $E_F\gtrsim1$eV, this must be extended to also include trigonal warping. More generally, the full hexagonal tight-binding model can be considered which is also suitable to treat chemical potentials around the van Hove singularity at $\sim3$eV. 

The polarizability of the full tight-binding model was discussed numerically\cite{Stauber10a} and within the semiclassical Boltzmann equation.\cite{Hanson13} Interestingly, one can also obtain analytical results for $\chi_{\rho\rho}$ for small q-vectors in the high-symmetry direction $\Gamma-M$.\cite{Stauber10c} The analytical solution displays the characteristic square-root singularity at the one-particle dispersion $\omega=v_Fq$ independent of the doping-dependent Fermi velocity which becomes zero at the van Hove singularity. For general ${\bm q}$-direction, this singularity splits in two peaks and acoustic plasmons were predicted due to different group velocities.\cite{Pisarra13}

Large Fermi energies up to $E_F=1.5$eV are, e.g., realized in intercalated graphene.\cite{Khrapach12} But the inclusion of lattice effects has only little effect on the low-frequency plasmon dispersion with an induced anisotropy within 1\%. Nevertheless, at energies close to $\hbar\omega\sim3$eV, i.e., the van Hove singularity, a linear dispersing damped plasmon mode emerges due to interband transitions.\cite{Hill09} Also for large wave vectors close to the corners of the hexagonal Brillouin zone, new low-frequency plasmon modes with a linear spectrum, so-called ÒintervalleyÓ plasmons, emerge which are related to the transitions between the two nonequivalent Dirac cones.\cite{Tudorovskiy10}

\subsubsection{Acoustic intraband plasmons}
\label{Sec:AcousticPlasmons}
Apart from the above mentioned acoustic plasmons due to interband or intervalley scattering, the optical $\sqrt{q}$-plasmons can also be converted into charged acoustic (intraband) plasmons. This is due to the strong screening of a metallic gate \cite{Principi11} or of a substrate with a huge dielectric constant.\cite{Stauber12} The sound velocity characterizing the acoustic plasmon dispersion $\omega=v_sq$, then reads
\begin{equation}
v_s=\sqrt{4\alpha_gk_Fz}v_F\;,
\end{equation}
with $z$ the distance of the graphene layer to the metallic gate or substrate. This approximation breaks down for small $k_Fz$ since the sound velocity cannot become smaller than the Fermi velocity and a more careful analysis is needed.\cite{StauberPRB12,Profumo12} Linear collective dispersions are also found from a general analysis of the plasmon spectrum of graphene in the vicinity of a thick plasma-like substrate.\cite{Horing09}

\subsection{Gapped Dirac Fermions}
The spectrum of graphene on various substrates like Boron Nitride\cite{Hunt13} or Iridium\cite{Rusponi10} shows a one-particle gap.
Gapped Dirac Fermions can also approximately describe a number of new 2D crystals like molybdenum disulphide, MoS${}_2$, or other transition metal dichalcogenides.\cite{Rostami13, Kormanyos13} 

For undoped, but gapped graphene, the response is similar to the case of doped, but ungapped graphene by identifying the gap parameter, $\Delta$, with twice the Fermi energy, $2E_F$. For instance, the local conductivity for neutral graphene with one-particle gap $\Delta$ reads
\begin{eqnarray}
\re\sigma&=&\sigma_0\frac{(\hbar\omega)^2+\Delta^2}{(\hbar\omega)^2}\theta(\hbar\omega-\Delta)\;,\\
\im\sigma&=&\frac{\sigma_0}{\pi}\left(\frac{2\Delta}{\hbar\omega}-\frac{(\hbar\omega)^2+\Delta^2}{(\hbar\omega)^2}\ln\left|\frac{\Delta+\hbar\omega}{\Delta-\hbar\omega}\right|\right)\;.
\end{eqnarray}
In case of large particle gaps, the non-relativistic limit is obtained.\cite{Scholz11} For the general, doped case, analytical results for the density response, $\chi_{\rho\rho}$,\cite{Pyatkovskiy09} as well as for the current response, $\chi_{jj}^{\pm}$,\cite{Scholz11} can be obtained and the corresponding plasmonic excitations were discussed within the RPA.\cite{Qaiumzadeh09} 

A gap in the one-particle spectrum can also be provoked artificially by graphene anti-dot lattices. In addition to the typical bulk plasmons in doped samples, also inter-band plasmons appear.\cite{Yuan13} These shall be discussed in detail in the next subsection.

\subsection{Interband plasmons and EELS}
\label{sec:AcousticPlasmon}
Before the technological advances to efficiently couple light to graphene by various near-field techniques, graphene plasmons were mainly investigated by means of high resolution electron energy-loss spectroscopy (EELS). Two regimes were discussed, i.e., acoustic interband plasmons at low and high energies, which will be addressed below.

\subsubsection{Acoustic plasmons at low energies}
For low energies $\hbar\omega\lesssim0.5$eV, EELS was first performed for graphene on SiC.\cite{Liu08} These experiments were repeated\cite{Langer10,Tegenkamp11} and extended to various metallic substrates like Platinum(111)\cite{Politano11} and Iridium(111).\cite{Langer11} 

All experiments in common is a characteristic peak in the loss function with linear dispersion at larger energies, even though the systems are quite different. E.g., graphene on SiC is doped with $E_F\approx0.3$eV and graphene on Iridium is undoped and gapped with $\Delta\approx0.1$eV. Additionally, the width of the resonances shows linear behavior in all cases. 

Both features, linear dispersion and linearly increasing line-width, can be captured by assuming the $q$-dependent conductivity of neutral graphene including only interband transitions, Eq. (\ref{ConductivityNeutral}). The loss function $S=-\im\epsilon_{RPA}^{-1}$ is then given by 
\begin{equation}
S(q,\omega)=\frac{x}{1+x^2}\;,\;\text{with }x=\frac{\pi\alpha_g}{2\epsilon}\frac{v_Fq}{\sqrt{\omega^2-(v_Fq)^2}}
 \end{equation}
which shows a maximum at $x=1$. This corresponds to a linear (acoustic) dispersion $\omega=v_sq$ with sound-velocity $v_s=\sqrt{1+\left(\frac{\pi\alpha_g}{2\epsilon}\right)^2}v_F$. For $\epsilon\approx3.5$, we obtain the experimentally observed sound velocity of $v_s\approx1.4v_F$ for a SiC-substrate.\cite{Liu08} 

The above analysis is practically unchanged, if we consider a lossy substrate with $\epsilon=\epsilon_R+\I\epsilon_I$ and the substitution $\epsilon\rightarrow|\epsilon|=\sqrt{\epsilon_R^2+\epsilon_I^2}$. For $|\epsilon|\approx3.5$, we obtain the experimentally observed sound velocity of $v_s\approx1.4v_F$  for Iridium.\cite{Langer11} For a Platinum substrate,  one finds $v_s\approx1.15v_F$ leading to a dielectric constant $|\epsilon|\approx6.1$.\cite{Politano11} 

\begin{figure}
\begin{center}
  \includegraphics[angle=0,width=0.495\columnwidth]{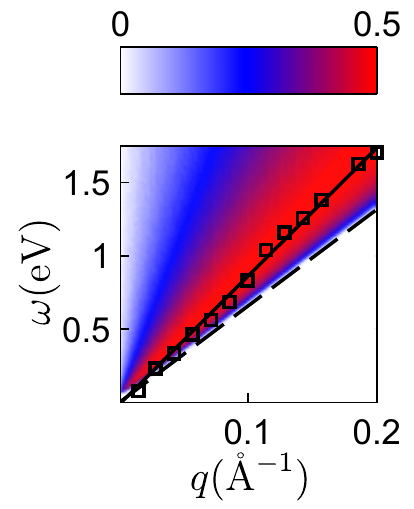}
  \includegraphics[angle=0,width=0.48\columnwidth]{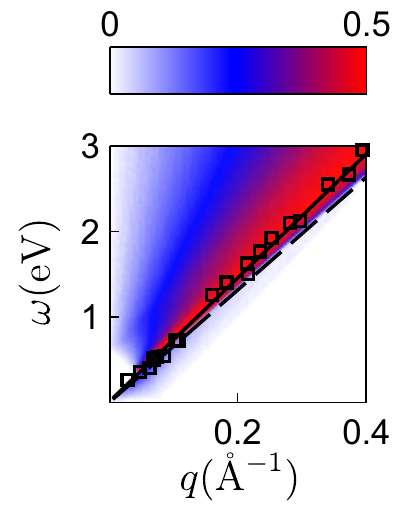}
\caption{(color online): Loss function $S(q,\omega)=-\im\epsilon_{RPA}^{-1}(q,\omega)$ of graphene on an Iridium (left) and Platinum (right) substrate  due to interband transitions compared to the experimental data of Ref. \onlinecite{Langer11} and \onlinecite{Politano11} (squares), respectively. Also shown the acoustic plasmon dispersion $\omega=v_sq$ with sound velocity $v_s\approx\sqrt{1+(\frac{\pi\alpha_g}{2|\epsilon|})^2}v_F$ for dielectric constants $|\epsilon|=3.5$ (left) and $|\epsilon|=6.1$ (right).} 
  \label{fig1}
\end{center}
\end{figure}

In Fig. \ref{fig1}, the electron loss function $S=-\im\epsilon_{RPA}^{-1}$ is shown together with the experimental data of Ref. \onlinecite{Langer11} (left) and Ref. \onlinecite{Politano11} (right). As mentioned above, a good fit for the linear dispersion $\omega=v_sq$ is obtained for $|\epsilon|\approx3.5$ (Ir) and $|\epsilon|\approx6.1$ (Pt) which lies considerably below the values of the corresponding local dielectric constants. The screening behavior of metals thus seems strongly reduced at finite $q$ which deserves further investigation. 

From the above analysis, it is clear that these "interband plasmons" are {\it no} collective excitations, but merely represent an enhanced charge resonance, i.e., they do not correspond to $\epsilon_{RPA}=0$. 

\subsubsection{Acoustic plasmons at high energies} 
For large energies $\hbar\omega\approx5$eV, a peak in the loss function associated to ${\bm\pi}\rightarrow{\bm\pi}^*$ transitions around the van Hove singularity was first predicted by DFT-studies,\cite{Kramberger08} and later experimentally observed in suspended graphene by electron energy-loss spectroscopy (EELS).\cite{Eberlein08} These $\pi$-plasmons also display a linear dispersion. Within the hexagonal tight-binding model and RPA, i.e., without including correlation or renormalization effects, no zero of the dielectric function $\epsilon_{RPA}$ is obtained.\cite{Stauber10a} The absorption peak would thus be merely due to interband transitions enhanced by a band-structure effect. Nevertheless, in bi- or multilayer, $\epsilon_{RPA}(\bm{q},\omega)$ becomes zero around the M-point and genuine plasmons emerge.\cite{Yuan11}

\subsection{Magneto-plasmons and strain}
A magnetic field strongly alters the response of the electrons and thus the plasmonic excitations.
The resulting magneto-plasmons have been studied, within different approaches, in Refs. \onlinecite{Iyengar07,Shizuya07,Roldan09}. They were observed in graphene epitaxially grown on SiC, where the Drude absorption is transformed into a strong terahertz plasmonic peak due to natural nanoscale inhomogeneities, such as substrate terraces and wrinkles.\cite{Crassee12} Similar experiments were also performed in a graphene disk array\cite{YanMagneto12} and graphene nanoribbons.\cite{Poumirol13}

It was further shown that the excitation of the plasmon modifies dramatically the magneto-optical response and in particular the Faraday rotation.\cite{Crassee12} The giant Faraday rotation due to magneto-plasmons in graphene micoribbons was also recently analyzed theoretically.\cite{Tymchenko13}

Due to the linear dispersion of Dirac Fermions, non-homogeneous strain and thus a variable hopping amplitude gives rise to pseudo-magnetic fields.\cite{Guinea10} The influence of strain on the response function was discussed in Ref. \onlinecite{Pellegrino11} and on plasmons in Ref. \onlinecite{Dugaev12}.

\subsection{Dissipative effects}
Intrinsic dissipation such as one-particle scattering or temperature naturally damp plasmonic excitations and limit the propagation length of the light-like density waves. Experimentally, the damping rate seems to be larger than what would be expected from the Drude formalism.\cite{Fei12} This was traced back to the large absorption plateau of gated graphene for energies below the absorption threshold, $\hbar\omega\lesssim2E_F$.\cite{Li08} Including impurity scattering due to short-range and Coulomb scatterers as well as electron-phonon interaction, the residual absorption could partially be explained,\cite{StauberPeres08,Peres09,Peres10} but important questions concerning the value of the plateau conductivity remain.\cite{Buljan13} Recently, electron-electron interactions were included to address these discrepancies,\cite{Principi13} and below, we will discuss this and additional lifetime limiting processes in more detail.

\subsubsection{Phonons}
Due to the low carbon mass, the energy of optical phonons of graphene is as large as 0.2eV. Below this energy threshold, there are no other prominent decay channels and for THz frequencies, long-lived plasmon excitations seem possible.\cite{Jablan09} But for large gate voltage, the plasmon dispersion hybridizes with the phonon modes which results in three new branches,\cite{Yan13} leading to plasmon lifetimes of 20fs or less when damping via the emission of graphene optical phonons is allowed. In Ref. \onlinecite{Fang13}, similar experiments with graphene nano disks have been performed,  yielding a larger lifetime approximately agreeing with the estimate coming from dc transport experiments.

Furthermore, surface polar phonons in the SiO${}_2$ substrate under graphene nanostructures lead to a significantly modified plasmon dispersion and damping, in contrast to the case of a nonpolar diamond-like-carbon substrate.\cite{Yan13} Surface phonons can be treated by using a frequency dependent dielectric function. For a polar substrate, it is usually parametrized by
\begin{equation}
\epsilon(\omega)=\epsilon_\infty\left(1+\frac{\omega_{LO}^2-\omega_{TO}^2}{\omega_{TO}^2-\omega(\omega+\I\gamma)}\right)\;,
\end{equation}
with the phonon frequencies $\omega_{LO}=1180$cm${}^{-1}$, $\omega_{TO}=1070$cm${}^{-1}$ and the damping rate $\hbar\gamma\approx1$meV in the case of SiO${}_2$.

\subsubsection{Electron-electron interaction}
The random-phase approximation is valid for wave numbers below the Thomas-Fermi screening length $\propto k_F$. Including vertex corrections in the bare charge response might lead to further dissipation channels, but the chiral nature of the Dirac carriers suppresses intrinsic plasmon losses when compared to parabolic band electrons in a 2DEG.\cite{Principi13}

\subsubsection{Nonlinear damping terms}
Another possible intrinsic damping mechanism is due to non-linear effects leading to an asymmetric broadening of the plasmon resonance.\cite{Mikhailov12} Mathematically, this was traced back to the singularity in the Boltzmann equation at the neutrality point. Following this reasoning, this dissipation should vanish in the case of gapped graphene, but the final expressions of Ref. \onlinecite{Mikhailov12} are independent of a mass-term. The effectiveness of this decay channel thus deserves more investigation, moreover, because this would question the general RPA-approach based on linear response.

\subsection{Beyond RPA}
The RPA has become a popular tool to analyze the screening and plasmonic properties of electronic systems mainly due to its simplicity. Obviously, it would be desirable to go beyond this first approximation by including more interaction terms which might have strong effects. 

For undoped graphene, vertex corrections were included in the bare polarizibility which leads to a novel plasmon mode in the region of intraband transitions.\cite{Gangadharaiah08,Sodemann12} For doped graphene sheets, a diagrammatic perturbation theory to first order in the electron-electron interaction was performed and proves that the plasmon frequency and Drude weight of the electron liquid might be  enhanced even in the long-wavelength limit.\cite{Abedinpour11} 

Alternatively, the $G_0W$-approximation is employed where the self-energy is calculated within the Born approximation based on the bare electronic Green function $G_0$ and the RPA-dressed photon Green function $W$.\cite{Polini08,Hwang08} With this approximation, angle resolved photoemission spectroscopy (ARPES) can be analyzed. APRES for epitaxially grown graphene, e.g., showed that interaction effects indeed lead to measurable changes in the energy spectrum.\cite{Bostwick07}  These changes can be interpreted in terms of new quasi-particles, so-called plasmarons, that arise due to the interaction between charge carriers and plasmons.\cite{Bostwick10}

\subsection{Plasmons in patterned graphene}
Plasmons cannot be directly excited by propagating electromagnetic radiation because the conservation of momentum is not satisfied in the photon absorption process. But periodically modulated sub-wavelength structures enable the direct coupling between propagating photonic modes and matter, see Fig. \ref{figure1}. For graphene, this has been achieved by a one-dimensional grating of nanoribbons,\cite{Ju11,Yan13,Strait13} and also photonic-crystal-like structures.\cite{Yan12,Fang13}   

The plasmon dispersion in quasi-one dimensional arrays depends on the width of the nanoribbon and the energy is lowered compared to 2D bulk plasmons due to the dipole-dipole interaction between the ribbons. The internal excitations in periodic structures have been investigated theoretically for photonic crystal-like structures based on disks\cite{Thongrattanasiri12}  and anti-dots,\cite{NikitinAPL12} nanoribbons,\cite{Nikitin12} and modulated nanowires.\cite{FerreiraPeres12,Peres12,Peres13} Also polarization-sensitive and gate-tunable photodetection in graphene nanoribbon arrays was demonstrated.\cite{Freitag13} For more details, we refer to the pedagogical review of Ref. \onlinecite{BludovIJMPB13}.

To numerically solve the Maxwell equations for general guided wave structures, 2D finite-difference time-domain (FDTD) or finite difference frequency domain (FDFD) techniques are widely used.\cite{Yee66,Zhao02} Since retardation effects can often be neglected, self-consistent eigenvalue equation combining graphene's response with the Poisson equation yield similar results.\cite{Christensen12,Strait13} For a linear stripe in $y$ direction, this can be formulated similarly to the hydrodynamic equation of Eq. (\ref{HD}):
\begin{eqnarray}
\rho(x)&=&\frac{\chi_{jj}^+}{\omega^2}(q^2-\partial_x^2)f(x)\nonumber\text{, with}\\
f(x)&=&\frac{1}{2\pi}\int dx' K_0(q|x-x'|)\rho(x')\;,
\end{eqnarray}
where $\rho$ is the charge density, $q$ the conserved momentum in $y$-direction and $K_0$ denotes the modified Bessel function of the third kind. The above set of equation has to be solved self-consistently.

\section{Plasmons in bilayer graphene}
When exfoliating graphene by micromechanical cleavage (scotch tape) techniques, one naturally produces graphene flakes with various number of layers $N$. These layers are normally Bernal or AB stacked and the unit cell contains $2N$ atoms in which half of all sites are vertically aligned. These multilayer graphene systems can be well described by including a interlayer hopping term changing the electronic spectrum and response. In this section, we will limit ourselves to bilayer graphene, $N=2$ and discuss plasmonic excitations in AB-stacked, AA-stacked and also twisted bilayer graphene where the two layers are rotated with respect to an arbitrary angle.  

\subsection{Minimal stacked bilayer graphene}
At low energies, Bernal stacked bilayer can be described by two parabolic bands touching at the Dirac points, leading to a Berry phase of 2$\pi$.\cite{McCannFalko06} The plasmonic spectrum shows a transition from Dirac to 2DEG plasmons,\cite{Wang07} and analytical formulas for the effective parabolic two-band model were first presented in Ref. \onlinecite{Sensarma10}. This was also recently discussed for finite temperature.\cite{DasSarma13}

The full tight-binding model possesses four bands, including also the corresponding anti-bonding modes. Analytical formulas for the four-band model were given in Ref. \onlinecite{Gamayun11,Yuan11} and similar expressions can be found for graphene with spin-orbit coupling.\cite{Scholz12} The simultaneous treatment of Coulomb interaction between and inside the layer was discussed in Ref. \onlinecite{Borghi09} where the formalism is naturally based on in-phase and out-of-phase excitations (see also Sec. \ref{sec:DoubleLayer}). Let us finally note that due to an optically active phonon mode and a resonant interband transition at infrared frequencies, the plasmonic properties of bilayer graphene can be strongly modified, leading to Fano-type resonances, giant plasmonic enhancement of infrared phonon absorption and a narrow window of optical transparency.\cite{Yan13b,Low13} 

The spectrum of bilayer graphene can become gapped by breaking the inversion symmetry between the two layers, e.g., by applying an interlayer bias.\cite{McCann06}
%,Castro07} 
The dispersion relation is then given by a Mexican hat dispersion and even though the ground state is still a Fermi liquid, the response is anomalous for small, but finite energies due to the diverging density of states at the band edge.\cite{Stauber07} This leads to novel plasmonic modes, present even for undoped biased bilayer graphene, but the physical origin of these genuine interband plasmons remains to be elucidated.\cite{Chuang13}

Apart from Bernal or AB-stacked graphene, also AA-stacked graphene can be obtained from folded graphene or twisted bilayer graphene with very small twist angle. In this configuration, all atoms are vertically aligned leading to a Fermi ring rather than a Dirac point at neutrality. The plasmon modes were discussed in Ref. \onlinecite{Roldan13} and have the curious property of being independent of the chemical potential in the energy region in which the two Dirac cones cross.
\subsection{Twisted bilayer graphene}
\begin{figure}
   \centering
   \includegraphics[width=1.1\columnwidth]{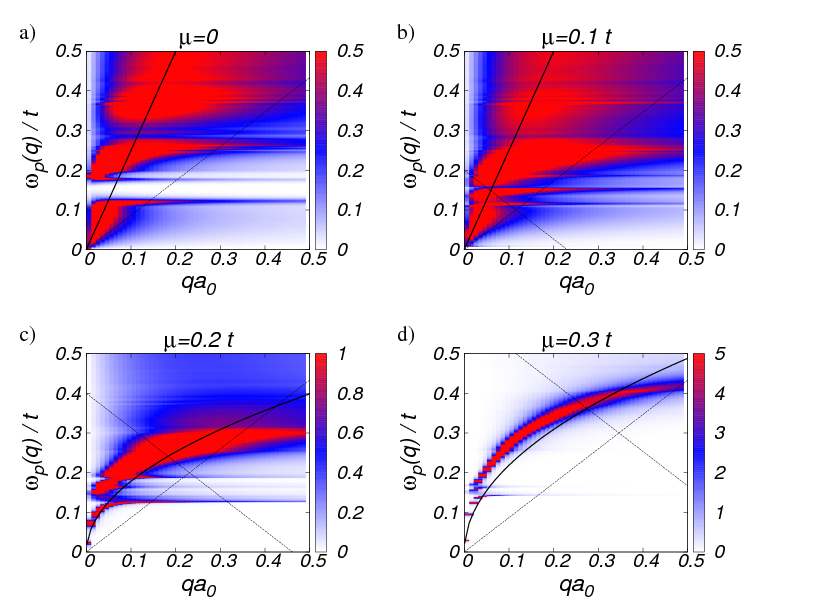} 
      \caption{Loss function $S(q,\omega)=-\im \epsilon_{RPA}^{-1}(q,\omega)$ in the long-wavelength RPA for twist angle $\theta=3.15^\circ$ and various chemical potentials $\mu/t=0,0.1,0.2,0.3,0.4$ with $\epsilon=2.4$ assuming a SiO${}_2$ substrate. The straight full line in a) and b) corresponds to acoustic interband plasmons, the curved full line in c) and d) to undamped intraband plasmons - both for the decoupled bilayer. The region of undamped plasmons of the decoupled bilayer is defined by thin dashed lines.}
   \label{fig:plasmons}
\end{figure}

Apart from minimal stacked AB or AA bilayer graphene, there is also turbostratic (twisted) graphene naturally obtained from epitaxially graphene grown on the carbon-terminated face of SiC. But even with mechanical cleavage techniques, these 2D carbon systems with internal rotational disorder can be produced and transferred to virtually any substrate, e.g. BN. 
 
For each valley, the electronic structure of twisted bilayer is defined by two Dirac points which are symmetrically separated by $\Delta K=2|K|\sin(\theta/2)$, $\theta$ being the twist angle and $|K|$ the modulus of the two $K$-points.\cite{Santos07} 
The electronic spectrum is characterized by a van Hove singularity located in between the two Dirac points at energy $\epsilon_M\approx\hbar v_F\Delta K/2$,\cite{Li10} which is repeated at higher energies due to an approximate shell-structure.

The plasmon dispersion can be discussed numerically based on the local dielectric function of Eq. (\ref{TwoDDielectric}) by first calculating $\re\sigma$,\cite{Moon13} and  then $\im\sigma$ by a subsequent Kramers-Kronig transformation.\cite{Stauber13} This gives rise to four possible plasmonic modes or resonances. i) There are undamped (conventional) graphene plasmons for chemical potentials with $\mu\ll\epsilon_M$ for which $\re\sigma=0$ and which are governed by twice the Drude weight of single layer graphene, $D=2D_{Dirac}$. This energy window becomes smaller for decreasing twist angle since the van Hove singularity moves closer to the neutrality point and is only relevant for large twist angles. ii) Due to the existence of several van Hove singularities, there are also interband "plasmons", see Sec. \ref{sec:AcousticPlasmon}. These are especially dominate for large twist angle and low $\mu$ and can lead to a broad optical gap in the interband excitations where $\im\sigma<0$, see Fig. \ref{fig:plasmons}a). iii) In the regime of large chemical potential, $\mu\gg\epsilon_M$, the conventional intraband plasmon is recovered, albeit Moir\'e-damped due to intrinsic (twist) disorder. The dispersion only depends slightly on the twist angle and becomes well-defined for large $\mu$, extending into the Landau damped region just as for the monolayer, see Fig. \ref{fig:plasmons}d). iv) Finally, due to the van Hove singularities, the imaginary part can become negative, $\im\sigma<0$, opening up the existence of transverse plasmons, see Sec. \ref{sec:Transverse}.   

In Fig. \ref{fig:plasmons}, the loss function, $S=-\im\epsilon_{RPA}^{-1}$, is shown for twist angle $\theta=3.15^\circ$ and various chemical potential with $\epsilon=2.4$ assuming a SiO${}_2$-substrate. Similar results are obtained for smaller angles.\cite{Stauber13} In all cases, the Dirac cone dispersion $\omega=v_Fq$ and $\hbar\omega=2E_F-\hbar v_Fq$ are shown as dashed lines, indicating the onset of intra- and interband transitions (see also Fig. \ref{fig:Diagram}). Also shown are the acoustic interband "plasmons" (see Sec. \ref{sec:AcousticPlasmon}) with sound velocity $v_s\approx\pi\alpha_gv_F/\epsilon$ (solid line in a) and b)) and the optical $\sqrt{q}$ plasmonic mode of   Eq. (\ref{TwoDPlasmons}) (solid line in c) and d)) for decoupled bilayer.
 
\section{Plasmons in general layered structures}
There is renewed focus on layered structures due to experimental advances in exfoliating a number of 2D materials and combining them in vertical stacks. Double-layer structures can thus be fabricated with relatively narrow and low energy barriers,\cite{Kim11} leading to novel devices like broadband optical modulator \cite{Liu11} or vertical field-effect transistors.\cite{Britnell12} 
%with boron-nitride \cite{Britnell12} or WS${}_2$ \cite{Georgiou13} buffer layer. 
%Also, the Coulomb drag of closely separated graphene layers was observed, which also depends on its dielectric response.\cite{Katsnelson11}  

Here, we present the basic steps how to derive the plasmon dispersion of multi-layer graphene or other 2D electronic systems including full retardation. We further assume that the layers only interact among themselves via Coulomb interaction; to also include coherent interlayer hopping, the 2D response function must be written in matrix form and would contain non-diagonal entries.\cite{Borghi09,Chuang13} An alternative method based on a simple analytical transfer-matrix approach can be found in Ref. \onlinecite{Kaipa12}.

\subsection{Undamped plasmons}
\begin{figure}[t] 
\begin{center}
  \includegraphics[angle=0,width=\linewidth]{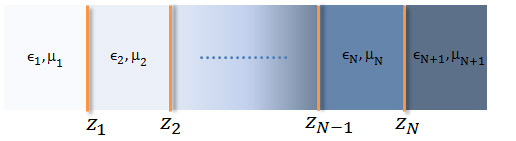}
\caption{(color online): Schematic setup of the multilayer graphene structure. The graphene (2DEG) layers, characterized by graphene's (2DEG) current response, $\chi_{jj}^{\pm,i}$, at position $z_i$, are surrounded by different dielectric media characterized by the relative dielectric constants $\epsilon_i$ and the relative magnetic permeabilities $\mu_i$.} 
  \label{figureMultiLayer}
\end{center}
\end{figure}

To discuss electromagnetic bound states, it is convenient to work within the Weyl gauge, setting the electrostatic potential equal to zero, $\phi=0$. We then only have to consider the vector field ${\bf A}$ and the coupling to a 2D electronic system (in the following, we will mainly discuss graphene) is entirely described by the current-current correlation function. We will treat the general multi-layer system of Fig. \ref{figureMultiLayer} and first discuss the longitudinal or $p$-polarization.

\subsubsection{Longitudinal or $p$-polarization}
\label{RetardedMultiLayer}
For longitudinal polarization, the general vector field has a component parallel and normal ($z$) to the interface,
\begin{equation}
{\bf A}({\bf r},z)=\sum_\q  e^{\I \q\cdot{\bf r}}\left(A^\parallel(\q,z){\bf e}_\q+A^\perp(\q,z){\bf e}_z\right)\;.
\end{equation}
The components of $A^\perp$ can be obtained from the components of $A^\parallel$ via the condition for a transverse field $\nabla\cdot{\bf A}=0$. It thus suffices to discuss the parallel component which is continuous at the interfaces.  With $A^\parallel=\sum_iA_i^\parallel$, we make the general ansatz for the gauge field in medium $i$,
\begin{equation}
\label{AnsatzAlong}
A_i^\parallel(\q,z)=a_ie^{-q_i'z}+b_ie^{q_i'z}\;,\;z_{i-1}\leq z<z_i\;,
\end{equation}
with the perpendicular wave vector $q_i^\prime=\sqrt{q^2-(\omega/c_i)^2}$ and $c_i=c/\sqrt{\epsilon_i\mu_i}$ the speed of light in the corresponding medium. The two boundary conditions at the $i$th interface are related to the continuity of the vector field and the discontinuity of the displacement field:
\begin{widetext}
\begin{eqnarray}
\label{BoundaryConditions}
a_{j}e^{-q_i'z_i}+b_ie^{q_i'z_i}&=&a_{i+1}e^{-q_{i+1}'z_i}+b_{i+1}e^{q_{i+1}'z_i}\\
q_{i+1}'(\epsilon_i-\alpha_i)a_{j}e^{-q_i'z_i}-q_{i+1}'(\epsilon_i+\alpha_i)b_ie^{q_i'z_i}&=&
\epsilon_{i+1}q_i'a_{i+1}e^{-q_{i+1}'z_i}-\epsilon_{i+1}q_i'b_{i+1}e^{q_{i+1}'z_i}
\end{eqnarray}
\end{widetext}
where $\alpha_i=e^2\chi_{jj}^{+,i}(q,\omega)\frac{q_i'}{\varepsilon_0\omega^2}$.

In the case of $N$ graphene interfaces ($z_0\to-\infty$, $z_{N+1}\to\infty$), we set $a_1=b_{N+1}=0$ and in the absence of dissipation ($\im\chi_{jj}^{+,i}=0$) we have a homogeneous set of $2N$ linear (real) equations with $2N$ variables, $Mx=0$. The condition det$M=0$ then yields $N$ plasmon modes with positive wavenumber $q$. In the non-retarded limit,\cite{Profumo10} they split into one optical mode with square-root dispersion and $N-1$ acoustic modes with linear dispersion for small $q$.\cite{Zhu13} 
 
For a single layer, the boundary conditions yield the implicit plasmon dispersion,
\begin{equation}
\label{PlasmonDisRet}
\omega^2 = e^2 \frac{q'_1q'_2}{\varepsilon_0(\epsilon_2 q'_1+ \epsilon_1 q'_2)} \chi_{jj}^+\;.
\end{equation} 
Neglecting retardation effects ($q_1'=q_2'=q$), and approximating the current response by the Drude weight $e^2\chi_{jj}^+\rightarrow D$, we recover the familiar expression for the plasmon dispersion $\omega_p^2=\frac{D}{\varepsilon_0(\epsilon_1+\epsilon_2)}q$. This provides the usually substitution rule for the dielectric constant in Eq. (\ref{TwoDPlasmons}), $\epsilon\rightarrow(\epsilon_1+\epsilon_2)/2$. The electron-electron interaction of 2D electrons is thus equally mediated through the upper and lower dielectric medium. 
\subsubsection{Transverse or $s$-polarization}
For transverse polarized light, only the parallel component is non-zero. We can thus write
\begin{equation}
\label{AnsatzA}
A^\parallel(\r,z)=\sum_\q e^{\I \q\cdot\r}A^\parallel(\q,z)
\end{equation}
and make the same ansatz as in Eq. (\ref{AnsatzAlong}):
\begin{equation}
\label{transverse}
A_i^\parallel(\q,z)=a_ie^{-q_i'z}+b_ie^{q_i'z}\;,\;z_{i-1}<z<z_i\;.
\end{equation}
 
The two boundary conditions at the $i$th interface are related to the continuity of the vector field and the discontinuity of the magnetic field. They are obtained from Eq. (\ref{BoundaryConditions}) by substituting  $q_i'\to \mu_i$, $\epsilon_i\to q_i'$, $\omega\to c$ and $\chi_{jj}^+\to-\chi_{jj}^-$.
\iffalse
\begin{widetext}
\begin{eqnarray}
a_{j}e^{-q_i'z_i}+b_ie^{q_i'z_i}&=&a_{i+1}e^{-q_{i+1}'z_i}+b_{i+1}e^{q_{i+1}'z_i}\\
(\mu_{i+1}q_{j}'-\alpha_i)a_{j}e^{-q_i'z_i}-(\mu_{i+1}q_{j}'+\alpha_i)b_ie^{q_i'z_i}&=&
\mu_iq_{i+1}'a_{i+1}e^{-q_{i+1}'z_i}-\mu_{j}q_{i+1}'b_{i+1}e^{q_{i+1}'z_i}
\end{eqnarray}
\end{widetext}
where $\alpha_i=\mu_0\chi_{t,j}^0(\q,\omega)$.
\fi
For a single layer, the plasmon dispersion is then defined by
\begin{equation}
\label{TEplasmons}
\mu_2q_1^\prime+\mu_1q_2^\prime+\mu_1\mu_2\mu_0e^2\chi_{jj}^-(\q,\omega)=0\;,
\end{equation}
whose possible solutions strongly depend on the surrounding dielectric media.\cite{Kotov13} For a discussion on general double layer graphene structures, see Ref. \onlinecite{Gutierrez13}.

Note that the symmetry between $p$- and $s$-polarization is normally not present. But here, we base our discussion on the parallel field component which is continuous in both cases, in contrary to the (usually discussed) total field which is (dis)continuous for transverse or $s$ (longitudinal or $p$) polarization. In the following, we will focus on the longitudinal polarization, but discuss in detail transverse plasmons in Sec. \ref{sec:Transverse}.

\subsection{Damped plasmons and energy loss function}
\label{EnergyLoss}
In the presence of dissipation, the plasmon dispersion ceases to be well-defined.
Therefore, to characterize damped plasmons, one frequently relies on the energy loss function defined as  $S(\bm{q},\omega)=-\im \epsilon^{-1}(\bm{q},\omega+\I0)$. This is a measure of the spectral density of the intrinsic plasmonic excitations: a sharp peak in $S(\bm{q},\omega)$ reveals long-lifetime plasmons; undamped plasmons, defined by $\epsilon(\bm{q},\omega)=0$, correspond to a delta peak in $S(\bm{q},\omega)$.
 
This formalism needs to be extended for two or more ($N$) interfaces and graphene's response is then given by a $N\times N$-matrix for each polarization (we will drop this index in the following). Within RPA, this gives the following matrix equation:
\iffalse
\begin{equation}\label{RPAchi}
\bm \chi_{RPA}^\pm = (\bm 1 -  e^2\bm \chi_{jj}^\pm \bm d^\pm)^{-1} \bm \chi_{jj}^\pm\equiv(\bm\epsilon^\pm)^{-1}\bm \chi_{jj}^\pm\;, 
\end{equation}  
%
For the special case of two graphene layers, the matrix $\bm \chi^\pm = \text{diag}(\chi_{jj}^{\pm,1},\chi_{jj}^{\pm,2})$ represents the bare  graphene's response in layer 1 ($\chi_{jj}^{\pm,1}$) and layer 2 ($\chi_{jj}^{\pm,2}$). $\bm d^\pm$ is the (bare) photon propagator in the absence of graphene ($\chi_{jj}^{\pm,i}=0$), but with the dielectric geometry (for $N$ layers, see Fig. \ref{figureMultiLayer}). The entries of the $N\times N$-matrix $\bm d$ can be obtained from the standard
 matching conditions or, equivalently, using multiple scattering formalism.
Diagonalizing the response matrix $\bm\chi_{RPA}^\pm$, one can discuss the elementary excitations of the full system, separately.
\fi
\begin{equation}\label{RPAchi}
\bm \chi_{RPA} = (\bm 1 -  e^2\bm \chi_{jj} \bm d)^{-1} \bm \chi_{jj}\equiv\bm\epsilon_{RPA}^{-1}\bm \chi_{jj}\;.
\end{equation}  
The $N\times N$-matrix $\bm \chi_{jj}$ denotes the (bare) graphene response which is diagonal in the absence of (coherent) interlayer coupling. The $N\times N$-matrix $\bm d$ is the (bare) photon propagator in the absence of graphene ($\chi_{jj}^{i}=0$), but with the dielectric geometry of Fig. \ref{figureMultiLayer}. The entries of $\bm d$ can be obtained from the standard matching conditions or, equivalently, using multiple scattering formalism. Diagonalizing the response matrix $\bm\chi_{RPA}$, one obtains the elementary excitations of the full system, i.e., in-phase and out-of-phase mode in the case of $N=2$.

Let us now define the energy loss function for arbitrary multi-layer structures. We emphasize this point because the plasmonic spectrum was frequently discussed by $S=-{\rm Im}\epsilon_{RPA}^{-1}$, where the (scalar) dielectric function was obtained by $\epsilon_{RPA}=\det\bm\epsilon_{RPA}$.\cite{Hwang09} But this "loss function" changes sign and can thus not be interpreted as a (positive definite) spectral density. Instead of the determinant, one rather needs to discuss the trace of the dielectric matrix.\cite{Chuang13} But graphene's excitations correspond to the imaginary part of the full response, $\bm \chi_{RPA}$, and the relative response of the several layers might differ. It is thus more appropriate to define the following generalization of the energy loss function:
\begin{equation}\label{loss}
S(\q,\omega)= -{\rm{Im Tr}}\bm \chi_{RPA}(\q,\omega+\I0)\;
\end{equation}  
Since $S(\q,\omega)$ is related to the imaginary part of a causal function, it is strictly positive and reveals the presence of the intrinsic excitations of the multi-layer system. It is further invariant with respect to unitary transformations between the several layers.
 
\subsection{Double layer}
\label{sec:DoubleLayerMain}
For the special case of two graphene layers, the above matrix $\bm \chi_{jj} = \text{diag}(\chi_{jj}^{1},\chi_{jj}^{2})$ represents the bare  graphene's response in layer 1 ($\chi_{jj}^{1}$) and layer 2 ($\chi_{jj}^{2}$). The photon propagator $\bm d$ for the two polarization can be found in Ref. \onlinecite{StauberPRB12}.

Undamped (longitudinal) plasmonic excitations are defined as usual by the zeros of the dielectric function $\det\bm\epsilon_{RPA}=0$. For two layers without retardation, this is often written in terms of the charge density response of the two layers, $\chi_{\rho\rho} ^{1/2}$:
\begin{equation}
\label{det}
(1-v_1\chi_{\rho\rho} ^1)( 1-v_2\chi_{\rho\rho} ^2) - v _{12} ^2 \chi_{\rho\rho} ^1 \chi_{\rho\rho} ^2=0\;,
\end{equation}
where $v_{1/2}(q)$ and $v_{12}(q)$ are the intra- and interlayer  Coulomb interaction, respectively. For different dielectric media on the left ($\epsilon_1$) , center ($\epsilon_{2}$) and right ($\epsilon_3$) and $z=z_2-z_1$ the distance between the two layers, the general expressions for the intra- and interlayer are given by\cite{Profumo12,Badalyan12,Stauber12} $v_{1/2}=[\cosh(qz)+(\epsilon_{3/1}/\epsilon_{2})\sinh(qz)]v_{12}(q)$ and $v_{12}=e^2\epsilon_{2}/(\varepsilon_0qN)$  with $N=\epsilon_{2}(\epsilon_1+\epsilon_3)\cosh(qz)+(\epsilon_1\epsilon_3+\epsilon_{2}^2)\sinh(qz)$.

Including retardation effects, we have to solve
\begin{eqnarray}
\label{DoubleLayerCondition}
&&(q_2'\epsilon_1+q_1'\epsilon_2-q_2'\alpha_1)(q_2'\epsilon_3+q_3'\epsilon_2-q_3'\alpha_2)\\\nonumber
&-&(q_2'\epsilon_1-q_1'\epsilon_2-q_2'\alpha_1)(q_2'\epsilon_3-q_3'\epsilon_2-q_3'\alpha_2)e^{-2q_2'z}=0\;,
\end{eqnarray}
where $\alpha_i=e^2\chi_{jj}^{+,i} q_i'/(\varepsilon_0\omega^2)$ and $z=z_2-z_1$ again the distance between the two layers. 

Below, we will discuss the two elementary modes of this system in more detail and also comment on near-field amplification at certain energies.
\subsubsection{Optical and acoustic modes}
\label{sec:DoubleLayer}
The plasmonic spectrum of double-layer graphene is characterized by an bonding and anti-bonding mode due to the electrostatic coupling between the two layers.\cite{Hwang09} In the case of longitudinal plasmons this leads to an ordinary (optical) 2D plasmon with $\sqrt{q}$-dispersion, but with larger energy since the charges of the two layers oscillate in phase. It also leads to a linear (acoustic) plasmon mode where the charges oscillate out of phase. In the case of transverse plasmons, there is no charge accumulation in the graphene layer and we find either one mode (for small layer separation) or two plasmon modes (for large layer separation).\cite{StauberPRB12}  

Usually, the plasmon dispersion is well separated from the light cone and we can set $q_1'=q_2'=q_3'=q$. Another approximation is given by the local response valid in the long-wavelength limit $qz\ll1$, i.e., replacing the current response by the corresponding Drude weight, $e^2\chi_{jj}^{+,i}=D_i$. The optical mode for $\omega\gg v_Fq\rightarrow0$ is then obtained as
\begin{equation}
\label{OpticalModeDoubleLayer}
\omega_{op}^2=g_sg_v\alpha_g v_F^2(k_F^1+k_F^2)\frac{q}{\epsilon_1+\epsilon_3}\;,
\end{equation}
with graphene's fine-structure constant $\alpha_g=\alpha\frac{c}{v_F}\approx2.2$.
The acoustic mode reads for $(k_F^1+k_F^2)z/\epsilon_2\gg1$
\begin{equation}
\omega_{ac}^2=g_sg_v\alpha_g v_F^2z\frac{k_F^1k_F^2}{(k_F^1+k_F^2)}\frac{q^2}{\epsilon_2}\;.
\end{equation}

The optical mode only depends on the sum of the outer dielectric media $\epsilon_1+\epsilon_3$ whereas the acoustic sound velocity only depends on the dielectric medium in the center, $\epsilon_2$. This is a general result because for the optical (in-phase) mode the interfaces have the same homogeneous charge density in the limit $qz\to0$, thus not polarizing the inner medium. For the acoustic (out-of-phase) mode in the same limit, there are opposite homogeneous charge densities on the two sheets just like for a capacitor which in turn does not polarize the surrounding media. 

For general parameters, the acoustic mode must be obtained in terms of a Laurent-Taylor expansion including the full expression of the response function.\cite{Santoro88} The square-root singularity of $\chi_{jj}^+(q,\omega)$ at $\omega=v_Fq$ then guarantees that the sound velocity is always greater than the Fermi velocity, $v_s>v_F$.\cite{Stauber12} The general analytical expression has been obtained by Profumo et al.\cite{Profumo12} 

The range of applicability of the analytical formula for the optical mode, Eq. (\ref{OpticalModeDoubleLayer}), depends on the relative value of $\epsilon_1,\epsilon_3$ with respect to $\epsilon_2$ and is only valid if they are of the same order. The general formula valid for $qz\ll1$ is given by
\begin{equation}
	\frac{\omega_+^2}{g_sg_v\alpha_gv_F^2q}=\frac{\epsilon_2(k_F^1+k_F^2)+qz(\epsilon_1k_F^2+\epsilon_3k_F^1)+\sqrt{R}}{2\left[\epsilon_2(\epsilon_1+\epsilon_3)+qz(\epsilon_2^2+\epsilon_1\epsilon_3)\right]}
\end{equation}
with $R=\epsilon_2^2(k_F^1+k_F^2)^2-2qz\epsilon_2(k_F^1-k_F^2)(\epsilon_1k_F^2-\epsilon_3k_F^1)+(qz)^2(\epsilon_1k_F^2-\epsilon_3k_F^1)^2$. For a 3D topological insulator with $\epsilon_2\gg\epsilon_1,\epsilon_3$ and in the case of equal densities $k_F^1=k_F^2=k_F$, this simplifies to 
\begin{equation}
\label{OpticalModeTwo}
\omega_+^2=\frac {2g_sg_v\alpha_g v_F^2k_Fq}{\epsilon_1+\epsilon_3}\left[1+\frac{qz\epsilon_2}{\epsilon_1+\epsilon_3}\right]^{-1}\;.
\end{equation}
In this case, Eq. (\ref{OpticalModeDoubleLayer}) is only valid for $qz\epsilon_2/(\epsilon_1+\epsilon_3)\ll1$.  

The left hand side of Fig. \ref{fig:plasmonsDouble} shows the energy loss function defined in Eq. (\ref{loss}) at zero temperature and the full red lines stand for $\delta$-like undamped excitations which merge into the Landau-damped regime of interband transitions. The right hand side shows the loss function at finite temperature $T=T_F/4$ with the Fermi temperature $T_F=E_F/k_B$. The black line indicates the plasmon dispersion at $T=0$ obtained by Eq. (\ref{det}) without dissipation, $\im\chi_{jj}^{+,i}=0$. The finite-temperature plasmonic resonances are slightly red-shifted compared to the $T=0$ dispersion as was the case in single layer graphene.
\iffalse
\begin{figure}
   \centering
   \includegraphics[height=0.49\columnwidth]{figPlasT0Double.pdf} 
   \includegraphics[height=0.49\columnwidth]{figPlasTDouble.pdf} 
      \caption{Generalized loss function $S(q,\omega)= -{\rm{Im Tr}}\bm \chi(q,\omega+\I0)$ for doped graphene in units of $E_F/\hbar^2$ at zero temperature (left) and at $T=T_F/4$ (right). The region of undamped plasmons of the decoupled bilayer at $T=0$ is defined by straight lines. The black curves on the right hand side correspond to $T=0$.}
   \label{fig:plasmonsDouble}
\end{figure}
\fi
\begin{figure}
   \centering
   \includegraphics[height=0.465\columnwidth]{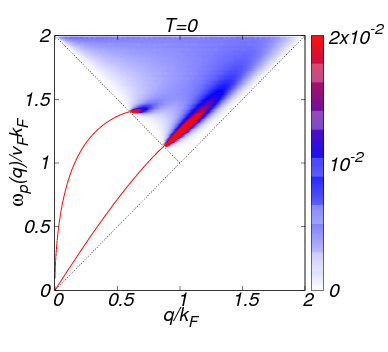} 
   \includegraphics[height=0.465\columnwidth]{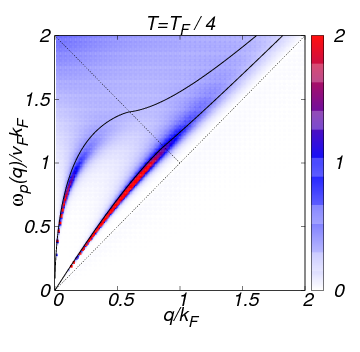} 
      \caption{Generalized loss function $S(q,\omega)= -{\rm{Im Tr}}\bm \chi_{RPA}(q,\omega+\I0)$ for doped graphene in units of $E_F/\hbar^2$ at zero temperature (left) and at $T=T_F/4$ (right). The region of undamped plasmons of the decoupled bilayer is defined by straight lines.}
   \label{fig:plasmonsDouble}
\end{figure}

\subsubsection{Near-field amplification}
In the case of two plasmon modes, there exists a frequency where the transmission is exponentially amplified, reminiscent to the situation of what happens in "Pendry's perfect lens".\cite{Pendry00} The frequency lies in between the two plasmon frequencies and is pinned to the out-of-phase mode for small wavenumbers. For large wavenumbers or interlayer distances, the two plasmon modes merge and sandwich the frequency of exponential amplification. The exponential transmission, 
\begin{equation}
T_{ex}=\frac{q_1'\epsilon_2-q_2'\epsilon_1+2q_2'\alpha_1}{q_3'\epsilon_2-q_2'\epsilon_3-2q_3'\alpha_2}\frac{q_3'}{q_1'}e^{(q_2'+q_3')z}\;,
\end{equation}
with $\alpha_i=e^2\chi_{jj}^{+,i} q_i'/(\varepsilon_0\omega^2)$ and $z=z_2-z_1$, is accompanied by zero reflection, $R_{ex}=0$, and similar expressions hold for transverse plasmons.\cite{StauberPRB12} 
For different densities in the two layers, the energy for near-field amplification depends on the arrangement of the layers. 

The possibility of exponential amplification might be useful for near-field microscopies and deserves further investigation for general multi-layer structures including dissipation.

\section{Plasmons including retardation effects}
Due to their large momentum, 2D longitudinal plasmons do not easily couple to propagating electromagnetic radiation and retardation effects can usually be neglected. But lowering the frequency, the unretarded square-root dispersion will finally cross the light-cone, pronouncing the onset of retardation effects, indicated by the circled region of Fig. \ref{figure1}a). 

On the other hand, in order to discuss transverse plasmons, full retardation is always needed since the plasmon dispersion is closely pinned to the light cone. In Sec. \ref{sec:Transverse}, we  will discuss general aspects of these excitations which lead to broadband polarization in graphene waveguides.\cite{Bao11} They are also present in gapped and one-dimensional structures as we will show below.
\subsection{Longitudinal or TM plasmons}
The standard expression for a 2D plasmon, $ \omega_p \propto \sqrt{q}
$, assumes instantaneous Coulomb coupling between
charges.\cite{Wunsch06,Hwang07} Therefore, it cannot be correct when
the nominal plasmon dispersion meets the light-cone, $\omega_p
\lesssim cq $. In this regime, even homogeneous graphene plasmons must couple strongly to (propagating) light and we will discuss the phenomena associated with the strong light-graphene coupling for single and double layer structures. 

\subsubsection{Single layer graphene}
We consider a single graphene sheet between two dielectrics with $\epsilon_1>\epsilon_2$. Graphene plasmons are then obtained from Eq. (\ref{PlasmonDisRet}) which includes retardation and in the unretarded limit, $c \rightarrow \infty $, this gives the  known
square-root dispersion. In contrast, the exact dispersion is linear
below a characteristic crossover frequency, $\omega_c$, and for reasonable dielectric constants, this scale is given by $\omega_c \sim \alpha \omega_F$  with $\omega_F=E_F/\hbar$. The plasmon dispersion thus merges with the light-cone of the slower medium and one obtains the following asymptotic behavior:\cite{GomezSantos12} 
\begin{equation}\label{dispersion} 
\left( \dfrac{\omega}{\omega_F}\right)^2 = \left \{ 
\begin{array}{ll}
\left(\frac{4 \alpha_g}{\epsilon_1 + \epsilon_2}\right) \left(\frac{q}{k_F}\right),  & \omega
\gtrsim \omega_c \\ \;&\\ \left(\frac{c_1}{v_F}\right)^2 \left(\frac{q}{k_F}\right)^2, & \omega
\lesssim \omega_c  
\end{array}
\right. 
\;,
\end{equation}    
with $c_1=c/\sqrt{\epsilon_1}$ the (slower) light velocity inside medium 1.
The crossover between the two regimes  takes place for frequencies  which roughly
corresponds to the intersection of the unretarded plasmon and light-cone
dispersion. This yields $\nu_c \sim 600
\,\text{GHz} $  for doping level  $n \sim 10^{13} \,\text{cm}^{-2}$, reaching
the technologically important  $\text{THz}$ regime  for  $n \sim 10^{14}
\,\text{cm}^{-2}$. 

The linear regime $\omega \lesssim \omega_c $ is also the region of strong graphene-light coupling. This can be seen by looking at the reflection and transmission amplitudes for the (in-plane) longitudinal
vector potential upon passing from medium $i$ to $j$, given by\cite{Stauber12} 
\begin{equation}\label{eqrij} 
r_{ij} = 
\frac{(\epsilon_i q'_j - \epsilon_j q'_i)\varepsilon_0\omega^2 + q'_i q'_j  e^2\chi_{jj}^+} {(\epsilon_i q'_j
+ \epsilon_j q'_i)\varepsilon_0\omega^2 -  q'_i q'_j  e^2\chi_{jj}^+}\;
,\end{equation}   
and $t_{ij}=1+r_{ij}$.

For interband transitions with $\omega\gtrsim2 \,\omega_F $, graphene's response
in Eq. (\ref{eqrij}) is small, leading to the universal $2.3 \%$ weak
absorption in vacuum.\cite{Mak08,Nair08} On the other hand, for $\omega\lesssim\omega_c $, graphene response starts to dominate in Eq. (\ref{eqrij}) implying
strong radiation-graphene coupling.  For instance, the   reflection amplitude
becomes $r \sim -1  $ for  $\omega \ll \omega_c $ , meaning (almost) perfect reflection
for  single-layer graphene. This perfect reflection is converted in perfect absorption when losses are allowed,\cite{Bludov10} providing a complementary and
potentially simpler alternative to absorption enhancement based on
periodic patterning.

\subsubsection{Double layer graphene}
Enhanced light-matter interaction also leads to {\it extraordinary transmission} for a double layer graphene arrangement. This term was originally coined to describe the enormous transmission experimentally
observed through periodically perforated metal
sheets, where naive
expectation would assume just the opposite.\cite{Ebbesen98} An explanation was
provided in terms of the excitation of  surface plasmons which results in 
enhanced (perfect without dissipative) transmission through a
nominally opaque region.\cite{GarciaVidal10} 

In the case of double-layer graphene where the central dielectric is less than the surrounding ones $\epsilon_2<\epsilon_1,\epsilon_3$, the resonant coherent excitations of the graphene layers also allow for the enhanced
transmission of photons through the central, classically forbidden region for
photons, in direct analogy with the metallic case.

The perfect transmission through the evanescent region is accompanied by a maximum in the spectral photonic density and thus due to a plasmonic response of the double layer graphene system. This enhanced light-matter interaction is also present in a general setup, in particular in the allowed (propagating) region where Fabry-P\'erot resonances emerge. These Fabry-P\'erot resonances become strongly quenched compared to the case without the graphene layers and the response, i.e., the spectral density displays a typical Fano-lineshape. We can interpret this as the formation of quasi-localized states between the doped graphene layers which slightly leak out and thus interact with the incoming (continuous) light field. 

The sharp response of graphene in
the absence of absorption also leads to enhanced absorption when losses are allowed and the setup is similar to the previously suggested enhanced absorption of graphene placed in a (double) Fabry-P\'erot cavity.\cite{Furchi12,Ferreira12} We finally note that there is a critical layer separation for the emergence of a separated acoustic mode, $z_c$. For layer separations with $z>z_c$, both modes, the in-phase and out-of-phase mode coincide with the light-cone.\cite{GomezSantos12}

\subsection{Transverse or TE Plasmons}
\label{sec:Transverse}
Collective charge density fluctuations are accompanied by collective longitudinal current fluctuations as dictated by the continuity equation. But there is also the possibility for collective transverse current fluctuations, see Eq. (\ref{chiTotal}). Whereas longitudinal plasmons can only exist for a metallic response, Re$\chi_{jj}^+>0$, transverse plasmons require a dielectric response, Re$\chi_{jj}^-<0$, because the photon propagator changes sign for the two polarization channels, see Eq. (\ref{dlt}). 

Transverse or TE plasmons are light-like excitations defined in Eq. (\ref{TEplasmons}) and were first discussed in Ref. \onlinecite{Ziegler07} in the case of suspended graphene. Due to their transverse nature, they are closely pinned to the light cone which makes them only weakly confined to the graphene sheet. For graphene on the interface of two distinct dielectric media, though, no solution is found because the different (local) light cones are too much separated in energy to simultaneously host the transverse plasmon.\cite{Kotov13} Also multi-layer structures with special dielectric media, e.g. superconductors with $\mu=0$, that might localize TE plasmons, do not lead to a solution after a critical layer separation.\cite{Gutierrez13}

In the retardation regime, the local optical conductivity is usually sufficient to discuss TE plasmons. This yields the following equation for suspended graphene:
\begin{equation}
1-\frac{\I\omega\sigma(\omega)}{c^2\sqrt{q^2-\omega^2/c^2}}=0
\label{TE}
\end{equation}
Writing the optical conductivity in terms of intra- and interband transitions, $\sigma=\sigma_{intra}+\sigma_{inter}$, with $\im\sigma_{intra}>0$, transverse plasmons can only be sustained by doped graphene when interband transitions with $\im\sigma_{inter}<0$ prevail. This is the case for energies $1.667<\hbar\omega/E_F<2$ where $\im\sigma<0$. With the dimensionless constants $Q=\hbar cq/E_F$ and $\Omega=\hbar\omega/E_F$, the TE plasmon dispersion in suspended graphene then reads\cite{Ziegler07} 
\begin{equation}
\label{TEPlasmonDispersion}
\sqrt{Q^2-\Omega^2}=2\alpha\left(\frac{\Omega}{4}\ln\left|\frac{2+\Omega}{2-\Omega}\right|-1\right)\;,
\end{equation}
where $\alpha$ is the fine-structure constant. 

Due to their small spectral weight, they have not been directly observed, yet. Nevertheless, strain will lead to field enhancement\cite{Pellegrino11} and also in bilayer graphene this mode is expected to be much stronger.\cite{Jablan11} In double layer structures with a certain layer separation, the TE mode can split in a bonding and anti-bonding mode and in between these two modes, a region of exponential near-field amplification can be defined.\cite{StauberPRB12} Another detection method is based on fluorescence quenching of a dye due to the presence of doped graphene and the non-radiative decay rate will entirely be defined by transverse plasmons at large distances.\cite{GomezSantos11}  

\subsubsection{Gapped graphene}
Without a charge density, one cannot generate longitudinal plasmons, i.e., collective charge density fluctuations. But it is possible to generate transverse plasmons, i.e., collective current fluctuations which are transverse to the direction of wave propagation. TE plasmons are thus present for gapped graphene even when the chemical potential lies inside the gap, $\Delta$. They are even more prominent compared to doped graphene ($2E_F\rightarrow\Delta$) due to the pure dielectric response of the system ($\sigma_{intra}=0$) and exist in the whole energy window $0<\hbar\omega/\Delta<1$. These collective current fluctuations associated to energies below the gap are possible since the generated particle-hole pairs are lowered in energy due to the attractive long-ranged Coulomb interaction. 

The dispersion of this mode, assuming local response, is pinned to the light cone but diverges logarithmically for energies close to the band gap in analogy to Eq. (\ref{TEPlasmonDispersion})
\begin{equation}
\sqrt{Q^2-\Omega^2}=\alpha\left(\frac{\Omega^2+1}{2\Omega}\ln\left|\frac{1+\Omega}{1-\Omega}\right|-1\right)\;,
\end{equation}
where $\alpha$ is the fine-structure constant and the dimensionless constants are $Q=\hbar cq/\Delta$ and $\Omega=\hbar \omega/\Delta$. For a gap of $\Delta\approx0.1$eV, the well-defined plasmonic modes would thus correspond to approximately 24Thz.

\subsubsection{One-dimensional transverse plasmons}
\label{Sec:TransversePlasmonsWire}
There can also exist purely transverse excitations in a quasi one-dimensional nanowire which is surprising since one would always expect charge accumulation at the border due to the transverse current oscillations. But for energies exponentially pinned to the light cone, the transverse component of the propagator becomes negative, opening up the possibility of transverse excitations for a dielectric response, see appendix.  

If we model the (graphene) nanoribbon by a cylindrical dielectric nanowire, characterized by a local susceptibility $\chi_e(\omega)>0$, the bare current response is given by $\chi_{jj}^-=-\omega^2\chi_e$. The self-consisting equation defining the current response then reads
\begin{equation}
\delta j^\alpha({\bf r})=-\chi_{jj}^-\left(A_{ext}^\alpha({\bf r})-\int d^2r'\mathcal{D}^{\alpha,\beta}({\bf r}-{\bf r}')\delta j^\beta({\bf r}')\right)\;,
\end{equation}
with $\mathcal{D}^{\alpha,\beta}$ the photonic propagator, see appendix. If the external field and the induced current distribution are homogeneous, we can average over the cylinder and obtain the following RPA-response function
\begin{equation}
\chi_{jj}^{RPA,-}=\frac{\chi_{jj}^-}{1-d_1^-\chi_{jj}^-}\;,
\end{equation}
with $d_1^-$ the one-dimensional transverse photon-propagator, see Eq. (\ref{App:TransProp}).

For a dielectric response $\chi_{jj}^-<0$, there are collective modes only if $d_1^-<0$. These modes are closely pinned to the light cone and with Eq. (\ref{App:TransProp}) we get
\begin{equation}
\left(q-\frac{\omega}{c}\right)\approx\frac{c}{2\omega a^2}\exp\left\{1-\left(1+\frac{2\pi a^2\epsilon_0}{\chi_e}\right)\frac{2c^2}{\omega^2a^2}\right\}\;,
\end{equation}
where $a$ is the radius of the cylinder.

We finally note that in three-dimensions, there are no collective transverse current oscillations which are separated from the light cone.

\section{Plasmons in Dirac systems with strong spin-orbit coupling}
The discovery of graphene triggered the search for other layered quasi-2D crystals like optically active 2D transition metal dichalcogenides MoS${}_2$, MoSe${}_2$ or WS${}_2$.
It also stimulated the search for new states of condensed matter resulting in a paradigmatic model for 2D topological insulators.\cite{Haldane88,Kane05a} Around the Dirac-points, it is given by the graphene Hamiltonian with a positive and negative gap with respect to the two $K$-points which arises from an intrinsic spin-orbit coupling. Including also Rashba spin-orbit coupling, one can tune the system from the quantum spin-Hall to the normal phase which are separated by a quantum critical point.  

Another system is represented by Hg(Cd)Te quantum wells\cite{Konig07} described by the Bernevig-Hughes-Zhang model.\cite{Bernevig06} In the quantum spin-Hall phase, electrons have intermediate properties between Dirac and Schr\"odinger fermiones which gives rise to plasmonic resonances even in the undoped limit.\cite{Juergens13}

In this section, we will discuss the plasmonic spectrum for graphene with intrinsic and Rashba spin-orbit coupling and the direct band gap 2D  semiconductor MoS${}_2$. We close with a discussion on the recently measured plasmonic spectrum of the 3D topological insulators Bi${}_2$Se${}_3$.\cite{DiPietro13}

\subsection{Graphene plasmons with spin-orbit coupling}
Let us define the graphene Hamiltonian with spin-orbit coupling in the Dirac cone approximation\cite{Kane05a}
\begin{equation}
H = v_F \vec p \cdot \vec \tau + \lambda_R \left( \vec \tau \times \vec \sigma \right) \vec e_z + \lambda_I \tau_z \sigma_z\;,
\label{Hamiltonian}
\end{equation}
where $\vec\sigma,\vec\tau$ denote the Pauli matrices referring to the pseudo-spin and valley degrees of freedom, respectively. For a sufficiently large intrinsic coupling parameter, $\lambda_I > \lambda_R$, the system is in the spin quantum Hall phase with a characteristic band gap.
For $\lambda_R > \lambda_I$ the gap in the spectrum is closed and the system behaves as an ordinary semi-metal.
At $\lambda_R = \lambda_I$ a quantum phase transition occurs in the system.

Closed analytical expressions for the complex polarizability of the above model have been obtained in Ref. \onlinecite{Scholz13} and we will follow this discussion. Similar to the case of massive Dirac Fermions,\cite{Pyatkovskiy09} and bilayer graphene,\cite{Gamayun11} this leads to several solutions of $\epsilon_{RPA}(q,\omega)=0$ for non-zero SOC parameters. One of these solutions has an almost linear dispersion with a sound velocity close to the Fermi velocity which exhibits an ending point for $\lambda_R\sim\lambda_I$ associated to a double zero of $\re\epsilon_{RPA}$. This solution does not lead to a resonance in the loss function and does thus not resemble a plasmonic mode. In the case where the gap in the spectrum is closed ($\lambda_R>\lambda_I$),
two additional zeros appear leading to potential high energy modes similar to bilayer graphene.\cite{Gamayun11,Yuan11} However, these potential collective modes are damped by interband transitions and no clear signature is seen in the density plot.

We are thus left with the solution corresponding to the genuine 2D plasmonic mode. Its dispersion $\omega_p$ can be approximated in the long-wavelength limit by 
\begin{equation}
\omega_p^2 = \frac{g_v\alpha_gv_F}{2\epsilon} \sum_{\nu=\pm1} \frac{k_{F\nu}^2}{\sqrt{k_{F\nu}^2 + \lambda_{-\nu}^2}}q\label{small_q_plasmons}\;,
\end{equation}
where $\lambda_\pm=\lambda\pm\lambda_I$ and the Fermi wave number $k_{F\pm}=\sqrt{\tilde\mu(\tilde\mu\mp2\lambda_R)\pm2\lambda_R\lambda_I-\lambda_I^2}$ with $\tilde\mu=\mu/(\hbar v_F)$.
\begin{figure}
\includegraphics[width=1.1\columnwidth]{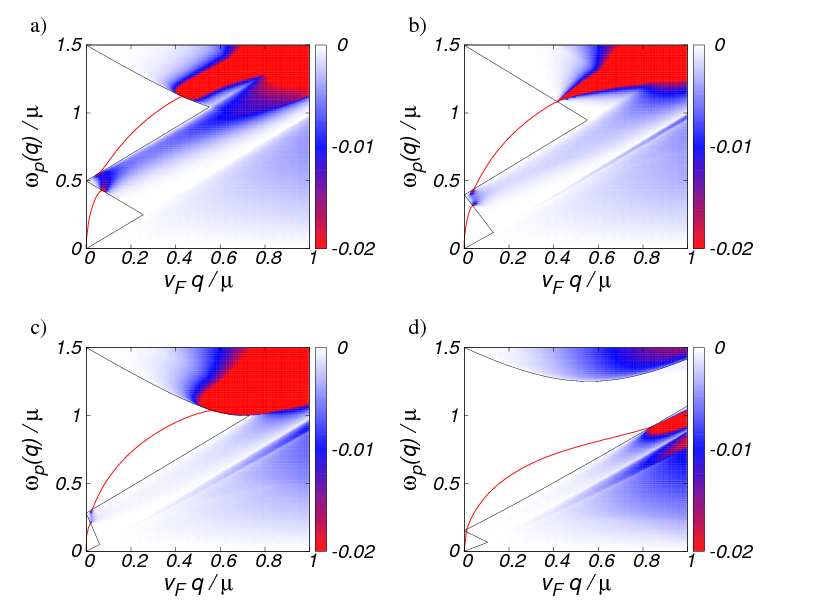}
\caption{Energy loss function  $S=-\im{\epsilon_{RPA}^{-1}(q,\omega+\I0)}$ of graphene with intrinsic $(\lambda_I)$ and Rashba $(\lambda_R)$ spin-orbit coupling for a) $(\lambda_R/\tilde\mu,\lambda_I/\tilde\mu)=(0.25,0)$, b) $(0.25,0.25)$, c) $(0.25,0.5)$, and d) $(0.25,0.75)$. The straight red lines show the undamped plasmon modes. The black lines indicate the boundaries of the particle hole continuum.
}
\label{FIG_spectral_weight}
\end{figure}

The numerical solution coincides with the long-wavelength solution for small momenta and then becomes red-shifted.  But for two occupied conduction bands, there is an additional Landau-damped region which is due to interband transitions from the two conduction bands and the plasmon mode is disrupted at $q\approx 0.05\tilde\mu$. At this  ``pseudo-gap'', the group velocity of the collective excitations formally diverges at the entering and exit points and the spectral weight is eventually transferred from the lower to the upper band as momentum is increased. 

The pseudo-gap of the plasmonic mode always emerges for $\lambda_R < 0.5\tilde\mu$, since the two bands are occupied independently of the value of $\lambda_I$, but it decreases for larger $\lambda_I$. This is shown in Fig. \ref{FIG_spectral_weight}, where the energy loss function for several values of $\lambda_{I,R}$ is plotted.

The spin-orbit coupling in graphene is small,\cite{Trauzettel06,Huertas06,Min06} still there are proposals how to enhance it,\cite{CastroNeto09} which might lead to interesting plasmonic systems when the above pseudo-gap region is reached.

\subsection{Plasmons in MoS${}_2$}
Around the corners of the Brillouin zone, a monolayer of MoS$_2$ can be described  
by an effective two-band model for both spin ($s = \pm 1$) and valley ($\tau = \pm1$) components.
\iffalse
\begin{eqnarray}
\hat H^{\tau s} &=& \frac \Delta 2 \sigma_z + \tau s \lambda \frac{1-\sigma_z}{2} + t_0 a_0 \vec k \cdot \vec\sigma_\tau \notag \\
&+& \frac {\hbar^2 k^2}{4m_0} \left( \alpha + \beta \sigma_z \right) 
+ t_1 a_0^2 \vec k \cdot \vec\sigma_\tau^* \sigma_x \vec k \cdot \vec\sigma_\tau^* .
\label{Hamiltonian}
\end{eqnarray}
\fi
A large energy gap of $\Delta = 1.9$eV separates the valence and 
conduction bands and the Hamiltonian for MoS${}_2$ is similar to the one of gapped graphene. The effective model of Refs. \onlinecite{Rostami13, Kormanyos13} also contains quadratic terms in the momentum which lead to different electron and hole masses. The plasmon dispersion in the long wavelength limit again displays the typical 2D behavior
\begin{equation}
(\hbar\omega_p)^2 = \frac{e^2}{8\pi \epsilon_0\epsilon}
\sum_{\tau,s=\pm1} k_F^{\tau\cdot s} \left|\frac{\partial E_\pm^{\tau s}}{\partial k}\right|_{k = k_F^{\tau\cdot s}} q \;,
\label{long_wavelength_plasmon}
\end{equation}
with the energy dispersion $E_\pm^{\tau s}$ for valley $\tau$ and spin $s$ given in Ref. \onlinecite{Scholz13} where the upper (lower) sign stands for the conduction (valence) band.  
\begin{figure}[t]
\includegraphics[width=0.495\columnwidth]{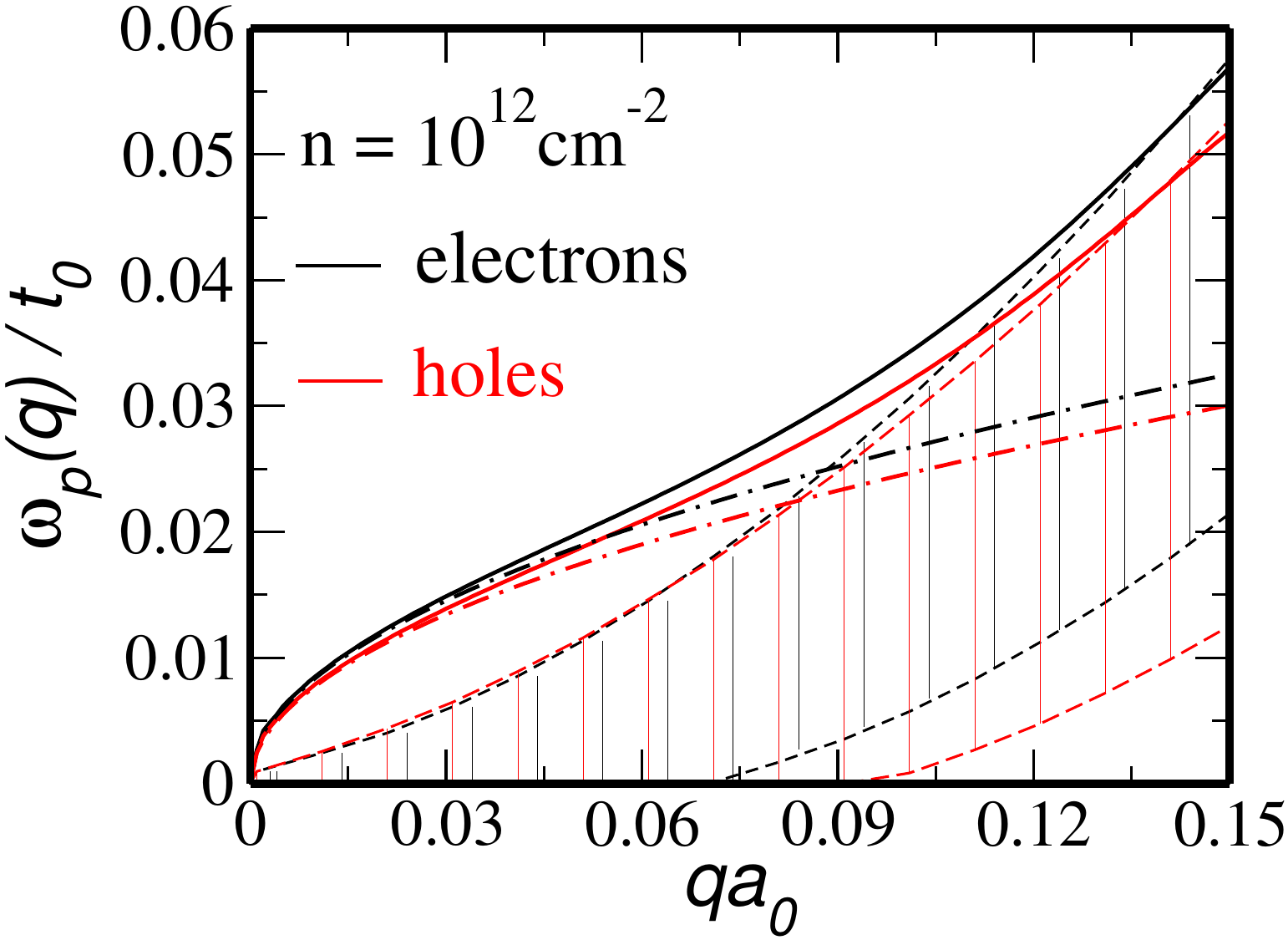}
\includegraphics[width=0.48\columnwidth]{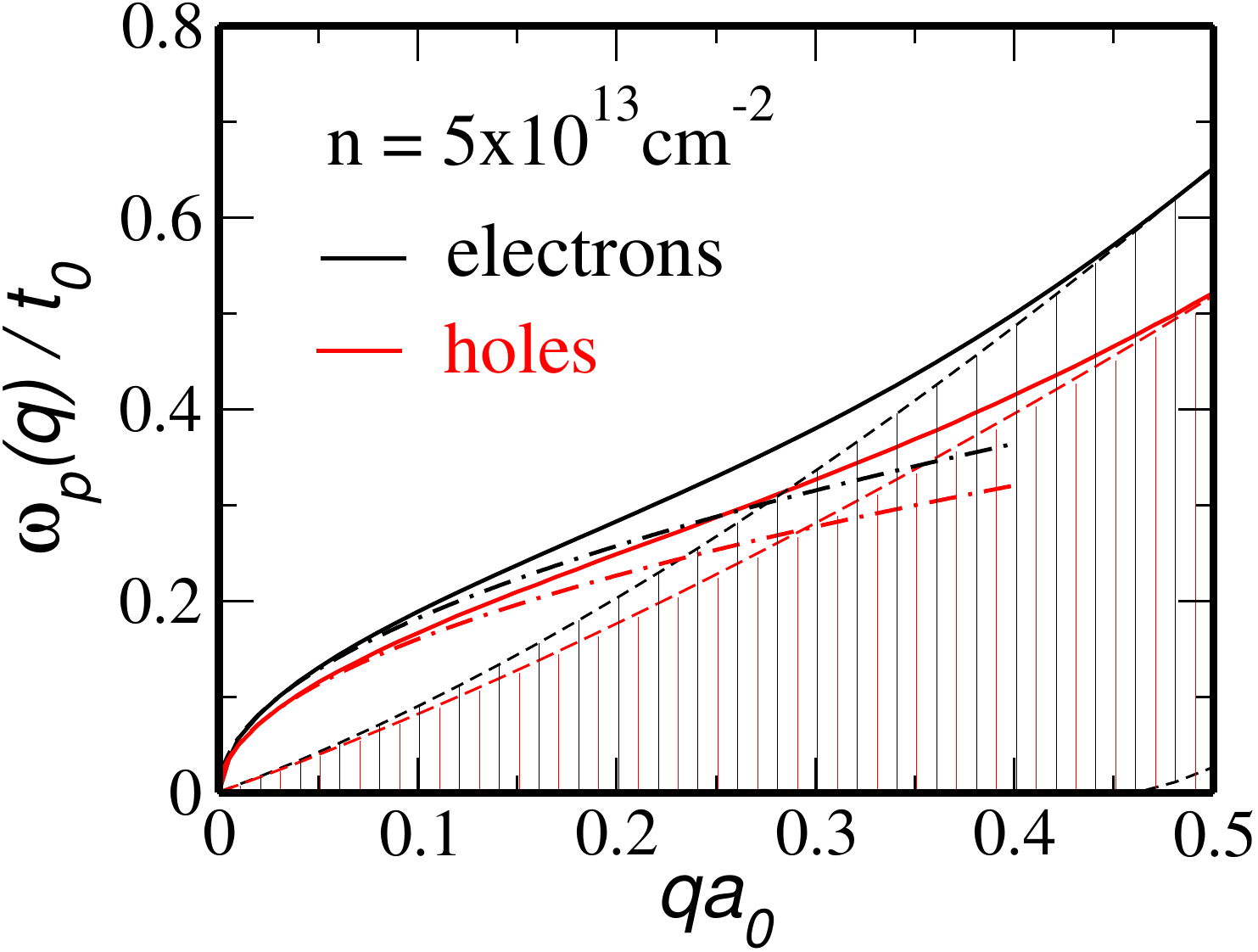}
\caption{(Color online) Plasmon dispersion of MoS${}_2$ for electron (black full line)
and hole (red full line) concentration of $n = 10^{12}$cm$^{-2}$ (left) and $n =5 \times 10^{13}$cm$^{-2}$ (right) in units of $t_0=1.68$eV and $a_0=0.184$nm. The dashed lines show the boundaries of the electron-hole continuum.
The dotted-dashed lines are the long wavelength results of Eq.~(\ref{long_wavelength_plasmon}). 
}
\label{FIG_Plasmons_low_doping}
\end{figure}

Due to the electron-hole symmetry in graphene,
plasmons in $n$- and $p$-doped samples show the same dynamics for equal
carrier concentrations.
This is no longer true in monolayer MoS${}_2$
as valence and conduction bands differ due to the strong spin-orbit coupling.
On the left hand side of Fig.~\ref{FIG_Plasmons_low_doping},
the plasmon dispersion and the intraband part of the electron-hole continuum are shown
at carrier concentration
$n =10^{12}$cm$^{-2}$
for electron (black) and hole (red) doping.
The dotted-dashed lines resemble the long-wavelength result of Eq.~(\ref{long_wavelength_plasmon}) and are in good agreement for $a_0 q \le 0.05$ ($a_0=0.184$nm).  The plasmon dispersions and the electron-hole continuum for $n$ and $p$ doping clearly differ and is enhanced for larger carrier density $n = 5\times10^{13}$cm$^{-2}$
as the difference in the electron and hole masses becomes more important, see right hand side of Fig.~\ref{FIG_Plasmons_low_doping}. 

Due to the large direct band gap in monolayer MoS$_2$, collective charge excitations enter the intraband electron hole continuum similar to 2D electron and hole gases with spin-orbit coupling.\cite{Pletyukhov06,Pletyukhov07,Badalyan09,Schliemann11,Kyrchenko11,Agarwal11,ScholzSchliemann13} The investigation of plasmonic effects in general transition metal dichalcogenides forms an active research area as they might determine the absorption and screening properties of single and multilayer systems.

\subsection{Plasmons in 3D topological insulators}
Typical 3D topological insulators (TI) like Bi${}_2$Se${}_3$ or Bi${}_2$Te${}_3$ are layered materials with repeating unit cells of hexagonal structure consisting of 5 layers. Due to the strong spin-orbit coupling, they display protected surface states that are characterized by a single Dirac cone whereas the bulk states show a full insulating gap. Dirac carriers at the surface of a TI reminds one of graphene, but in graphene, it is momentum and pseudo-spin that are constrained, whereas it is momentum and real spin which are locked in the case of these topologically protected edge states.\cite{Hasan10,Qi11} The collective modes of this "helical metal" were first discussed in Ref. \onlinecite{Raghu10} focusing on the curious fact that density fluctuations induce transverse spin fluctuations and vice versa.  Spin-plasmons were also discussed in terms of the plasmon wave function.\cite{Efimkin12b}  

\begin{figure}
\centering
  \includegraphics[height=0.5\columnwidth]{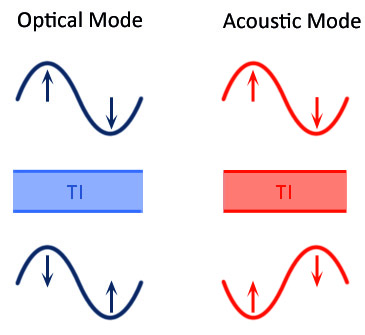}
  \includegraphics[height=0.5\columnwidth]{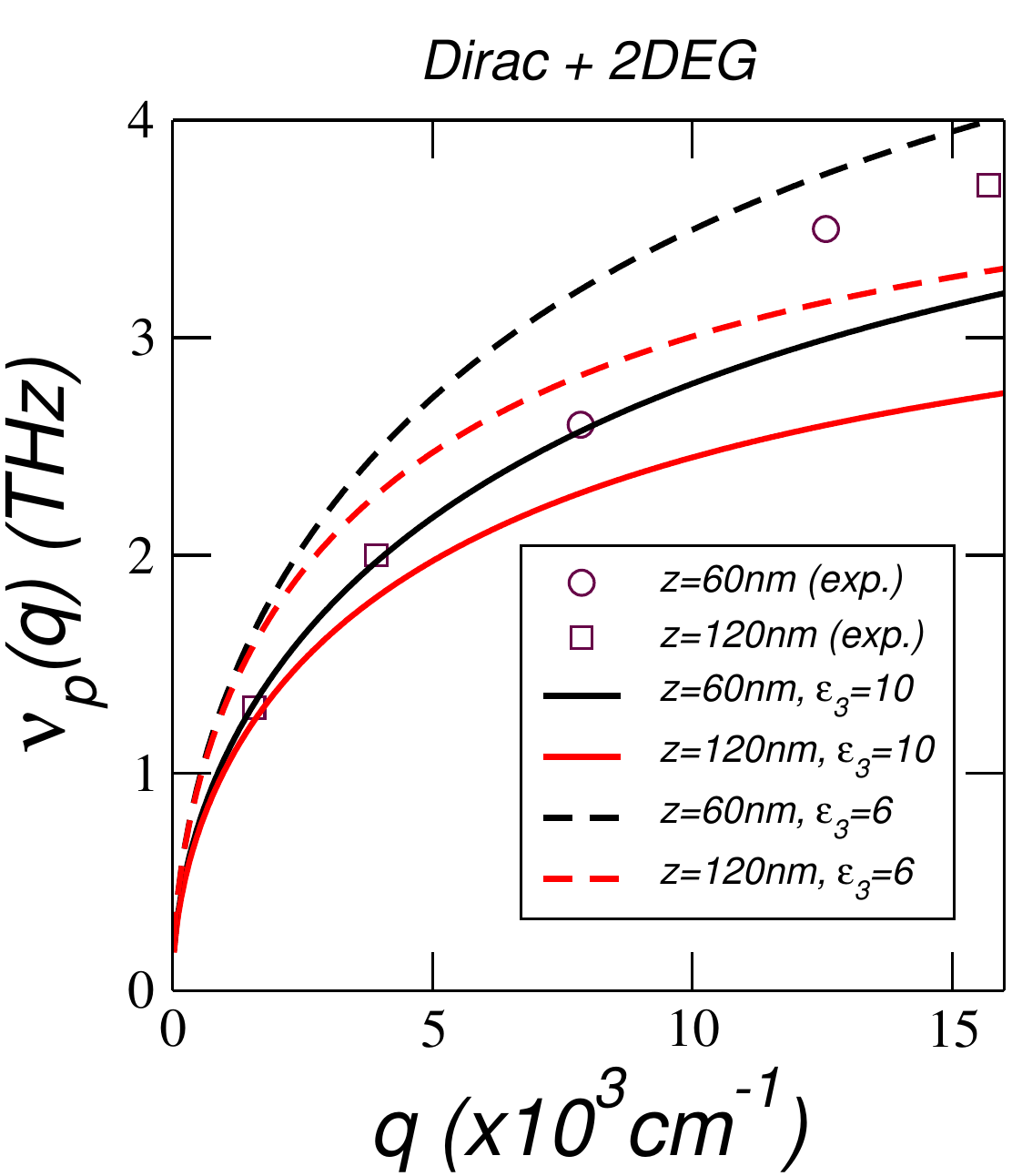}
\caption{(color online):  Left: Schematic picture of the spin-charge separation. For the optical mode, the charge waves are in-phase and the spin waves are in different directions for the top and bottom layer. This leads to effective (pure) charge oscillations. For the acoustic mode, the charge modulations are out-of-phase and the spin waves point in the same direction for the top and bottom layer. This leads to  effective (pure) spin oscillations. Right: Comparison of the experimental data of Ref. \onlinecite{DiPietro13} (symbols) with the optical mode of Eq. (\ref{OpticalModeTwo}) considering the response of Dirac Fermions and 2DEG for two slab width $z=60$nm (black) and $z=120$nm (red) for $\epsilon_3=10$ (full lines) and $\epsilon_3=6$ (dashed lines).}
  \label{SpinChargeSep}
\end{figure}
\subsubsection{Spin-charge separation}
Dirac cones must come in multiples of two and the single Dirac cone on one surface naturally finds its pair on the opposite side. A thin topological insulator slap without wave function overlap thus seems to mimic double-layer graphene (see Sec. \ref{sec:DoubleLayerMain}), because the charge response of a helical metal is identical to the charge response of graphene apart from a factor 4 (for TI $g_s=g_v=1$).  But the Dirac cone on one TI surface is {\it not} an identical copy of the Dirac cone on the other surface because the sign of the Fermi velocity must be opposite for the two Dirac cones. This means that the spin locked to the charge momentum is polarized in opposite directions on the two surfaces which has the curious consequence that in-phase and out-of-phase oscillations can be purely charge- and spin-like, respectively, see left hand side of Fig. \ref{SpinChargeSep}.\cite{StauberBrey13} 

Let us now consider the electronic motion confined by a quasi-one dimensional nanowire and assume that the ribbon with width $a$ is wide enough to justify the local approximation of the response function. Thus, only the in-plane Coulomb interaction needs to be modified, see appendix.
\iffalse
\begin{equation}
v_q^{1/2}=\frac{1}{2\pi\varepsilon_0}\ln\frac{\sqrt{e}}{qa}\;.
\end{equation}
\fi
The optical mode is then obtained as
\begin{equation}
\omega^2=(\alpha_d v_F^2k_Fa/\pi)q^2\ln(\sqrt{e}/(qa))\;. 
\end{equation}
Considering the logarithmic correction only as an additional factor, also the optical mode shows a linear dispersion.  

Spin-charge separation and collective excitations with linear dispersion are the characteristics of the Tomonaga-Luttinger phenomenology for one-dimensional electron systems. Choosing the width and the length of the ribbon as $a\approx100$nm and $L\approx 10\mu$m, we obtain $v_c\approx10v_F$ in the long-wavelength limit. For the sound (spin) velocity we set $v_s\approx v_F$, valid for small TI slab widths.This would correspond to one-dimensional interacting electrons with Luttinger liquid parameter $K\approx0.1$. By varying the ribbon width to $a\approx10$nm, one can reach $K\approx0.2$, thus being able to tune the effective interaction.  A helical Luttinger liquid in topological insulator nanowires was recently discussed in Ref. \onlinecite{Egger10} and offers a microscopic theory for the observed spectrum.

\subsubsection{Comparison to experiment}
Optical plasmon excitations have recently been detected using infrared spectroscopy as discussed in the introduction.\cite{DiPietro13} In order to explain their data based on a double-layer model, one not only needs to consider the charge response of the Dirac Fermions giving rise to a 4x4 response matrix including spin and charge channels, but also the 2DEG trapped underneath the TI surface, resulting in a 8x8 response matrix.\cite{Bansal12} Taking the depletion layer into account might thus change the number and behavior of plasmon modes, found in typical double layer structures.  

Nevertheless, due to the closeness of the depletion layer to the TI surface, the only newly emerging modes would be charge-less acoustic-like excitations formed by superpositions of the Dirac carriers and the depletion layer on the same (top or bottom) TI surface. These modes are thus closely pinned to the particle-hole continuum and not observable. They also do not affect the modes obtained by the initial 4x4 matrix which can further be reduced to a 2x2 matrix since only the charge channels are coupled. We can, therefore, set the effective response as $\chi=\chi^{Dirac}+\chi^{2DEG}$.\cite{StauberBrey13}

The full density response of a 2DEG was derived by Stern,\cite{Stern67} but here the local approximation is sufficient, i.e., Eq. (\ref{localCurrentResponse}) with $\nu=g_s=2$ and $g_v=1$. We can thus use Eq. (\ref{OpticalModeTwo}) with $\mu\to\mu^{Dirac}+4\mu^{2DEG}$ and with $\mu^{Dirac}=542$meV and $\mu^{2DEG}=60$meV,\cite{Bansal12} we obtain a reasonable fit to the experimental data for low wave numbers $q\lesssim10^4$cm${}^{-1}$. This can be seen on the right hand side of Fig. \ref{SpinChargeSep}, where we plot the resonant plasmon frequencies $\nu_p$ for slab widths $z=60$nm (black) and $z=120$nm (red) for a dielectric substrate with $\epsilon_3=10$ (full lines). We further assumed $\epsilon_1=1$ and $\epsilon_2=100$. 

The two high-energy plasmon resonances with $q>10^4$cm${}^{-1}$ cannot be well described by our fit and are blue shifted. This is in contrast with our expectations because the dipole-dipole interaction between the patterned nano wires should lead to an additional red-shift compared to the analytic curves for samples with small periodicities;\cite{Nikitin12,Christensen12}  and this shift can be as large as 20\%.\cite{Strait13}  A possible blue shift could be provided by including the frequency dependence of $\epsilon_B$ which might lead to smaller values $\epsilon_B(\omega)<10$. We, therefore, also show curves with $\epsilon_B=6$ (dashed lines) which value was measured for thin ($15$nm) Al${}_2$O${}_3$-films.\cite{Kim09} Also a decrease of $\epsilon_{TI}$ would lead to a blue shift for larger frequencies and further studies are needed to reconcile theory with experiment in this regime.

\section{Summary and Outlook}
In this topical review, we have presented and discussed various aspect related to plasmonic excitations in graphene nanostructures and other Dirac systems. Our discussion was based on linear response theory and the random phase approximation as well as on hydrodynamic approaches. The typical square-root dispersion of 2D electronic systems defined the plasmonic spectrum in most cases as suggested from phenomenological models. Still, retardation effects, strong screening and spin-orbit coupling can lead to a linear or disrupted spectrum. Also interband plasmons disperse linearly, even though these charge excitations do often not represent genuine plasmons defined by $\epsilon_{RPA}=0$, but are only manifested by a peak in the energy loss function. 

We also discussed plasmons in graphene-based heterostructures with inhomogeneous dielectric background. For intrinsic dissipation, we introduced the generalized loss function necessary to define the plasmonic multi-layer resonances. For double-layer structures, general analytical formulas in the long-wavelength limit were presented, especially important in the context of 3D topological insulators. Also retardation effects were discussed leading to enhanced absorption and quenched Fabry-P\'erot resonances, as well as to transverse plasmons even in undoped, gapped 2D and 1D materials.

%Several applications were not discussed. E.g., plasmons change the dipole-dipole interaction between emitter and graphene leading to non-radiative energy transfer and strong quenching.\cite{Huidobro12,Gaudreau13} 

%\cite{Huidobro12,Gaudreau13} 
Several aspects such as quantum effects for plasmons in quantum junctions of graphene dimers,\cite{Thongrattanasiri13} boundary effects giving rise to Mie-resonances in conducting nano particles,\cite{Zhao09} or fluorescent quenching\cite{GarciaVidal12,Koppens13} were not addressed. For a recent review focusing on the wide-ranged potential applications of graphene with respect to plasmonic metamaterials, light harvesting, THz technology, biotechnology or medical sciences, see Ref. \onlinecite{Luo13}. 

Another uncovered topic was the non-linear response of  graphene which is ten times larger compared to nobel metals like gold.\cite{Hendry10} This property can lead to an effective metamaterial with negative refractive index\cite{Harutyunyan13} or support the propagation of sub wavelength optical solitons.\cite{Nesterov13} Nonlinear plasmonics based on graphene thus promises to become an important research field in the future.\cite{KauranenNP12}
%efficiently generate THz plasmons in graphene and topological insulators (1308.2005) and can also confine transverse electric plasmon-polariton modes (Bludov).

\section{Acknowledgements}
We thank Nuno Peres, Paco Guinea, Luis Brey, \'Angel Guti\'errez, Rafa Rold\'an, John Schliemann, Andreas Scholz, Reza Asgari, F. J. Grac\'\i a Vidal, Javier Grac\'\i a de Abajo and especially Guillermo G\'omez-Santos for helpful discussions. This work has been supported by FCT under grant PTDC/FIS/101434/2008 and MICINN under grant FIS2010-21883-C02-02. 

 \section{Appendix: Linear Response}
The linear response  of a system to an external perturbation is related to correlation functions via the Kubo formula.\cite{Wen04,Giuliani05} For a general Hamiltonian $H=H_0+\delta H$ and $\delta H=\lambda^\alpha\psi^\alpha$, the response of a quantum field $\psi^\alpha$ shall be defined by $\delta\langle\psi^\alpha\rangle=\chi_{\psi\psi}^{\alpha,\beta}\lambda^\beta$ with
\begin{equation}
\label{LinearResponse}
\chi_{\psi\psi}^{\alpha,\beta}(\q,\omega)=-\frac{\I}{\hbar}\int_0^\infty e^{\I\omega t}\langle [\psi_\q^\alpha(t),\psi_{-\q}^\beta(0)]\rangle\;,
\end{equation}
where summation over repeated indices is implied. Depending on the context, we will call the response function retarded Green's function or propagator of $\psi^\alpha$. In the following, we will discuss the density, current and gauge field response. We will then point out the relation between the current and photon propagator. 

\iffalse
Before we summarize the linear response functions of the density, current for the general hexagonal tight-binding model as well as of the gauge field, it is noteworthy that the non-linear response of  graphene is ten times larger compared to nobel metals like gold.\cite{Hendry10} This property can lead to an effective metamaterial with negative refractive index,\cite{Harutyunyan13} or supports the propagation of sub wavelength optical solitons.\cite{Nesterov13} 
%efficiently generate THz plasmons in graphene and topological insulators (1308.2005) and can also confine transverse electric plasmon-polariton modes (Bludov). 
We will not cover this important aspect of graphene optics and refer to the original articles and Ref. \onlinecite{Kauranen12}.
\fi

\subsection{Density-response}
We first discuss the density-density correlation function or polarizability, i.e., $\psi^\alpha\to\rho$ is the electronic density operator and $\lambda^\alpha\to\phi$ the electrostatic potential. In 3D, this is also called the Lindhard function. For a 2D electron gas (2DEG), $\chi_{\rho\rho}$ was first calculated by Stern.\cite{Stern67} For a 2D Dirac system, it was first discussed by Shung in the context of intercalated graphite.\cite{Shung86}  

The 2D polarizability is sometimes defined including a minus-sign with respect to the convention in 3D, following the original work of Stern. Here, we will use the definition of Eq. (\ref{LinearResponse}), i.e., we have $\im\chi_{\rho\rho}<0$ for $\omega>0$ and the random phase approximation (RPA) is then defined as usual with a relative minus sign, see Eq. (\ref{RPAdensity}).
 
The density-density correlator or Lindhard function for the tight-binding model on a 2D (honeycomb) lattice is given by 
\begin{eqnarray}
\label{densitycorrelation}
\chi_{\rho\rho}(\q,\omega)&=&\frac{g_s}{(2\pi)^2}\int_{{\rm{1.BZ}}}d^2k\sum_{s,s'=\pm}f_{s\cdot s'}(\k,\q)\\\nonumber
&\times&\frac{n_F(E^{s}({\bf k}))-n_F(E^{s'}({\bf k}+{\bf q}))}{E^{s}({\bf k})-E^{s'}({\bf k}+{\bf q})+\hbar\omega+\I0}\;,
\end{eqnarray}
with the eigenenergies $E^{\pm}({\bf k})=\pm t|\phi(\k)|$ ($t\approx2.78$eV is the hopping amplitude), $n_F(E)=(e^{\beta(E-\mu)}+1)^{-1}$ the Fermi function, $g_s=2$ the spin-degeneracy and $\phi(\k)=\sum_{{\bm \delta}_i}e^{\I{\bm \delta}_i\cdot\k}$ the complex structure factor, with ${\bm \delta}_i$ the three nearest neighbor vectors of the hexagonal tight-binding model.\cite{Stauber10a} Due to the two gapless bands, the above expression contains the band-overlap function 
\begin{equation}
\label{overlapdensity}
f_{\pm}(\k,\q)=\frac{1}{2}\left(1\pm\re\left[\frac{\phi(\k)}{|\phi(\k)|}\frac{\phi^*(\k+\q)}{|\phi(\k+\q)|}\right]\right)\;,
\end{equation}
not present in the one-band 2DEG discussed by Stern. In the Dirac-cone approximation, the above integral can be solved analytically in terms of two analytic function.\cite{Wunsch06} We present this solution in the context of the longitudinal current response, see Eqs. (\ref{CurrentCurrent}) - (\ref{CurrentCurrentMu}). 

To discuss the plasmonic dispersion, the local approximation ($q\to0$) is frequently used which is given by\cite{Wunsch06,Hwang07}
\begin{equation}
\chi_{\rho\rho}(\omega)=\frac{g_sg_v q^2}{8\pi \hbar \omega}\Big[\frac{2\mu}{\hbar
\omega}+\frac{1}{2}\ln\left|\frac{2\mu-\hbar \omega}{2\mu+\hbar \omega}%
\right|  -i\frac{\pi}{2}\Theta(\hbar \omega-2\mu)\Big]\;,
\label{LongWave}
\end{equation}
with $g_v=2$ the valley-degeneracy. From this expression, we obtain the universal conductivity of undoped graphene $\sigma_0=\frac{\pi}{2}\frac{e^2}{h}$ via the continuity equation $\sigma=\I e^2\chi_{\rho\rho}\omega/q^2$.

Let us make a brief reminder on this universal conductivity. The local conductivity of graphene at zero temperature is not uniquely defined and can lead to two distinct universal expressions, i.e., both do not depend on any material constants, as was already noted in 1994.\cite{Ludwig94} Its value depends on whether the artificially introduced phenomenological damping term $\gamma\to0$ goes to zero before or after the (finite) frequency $\omega\to0$ and reflects in some way the duality of graphene being a (semi)metal with zero density of state or a semiconductor with zero band-gap. This ambiguity is not a mathematical artifact, but can be interpreted physically whether or not metallic leads give rise to a finite broadening in the dc-limit. In the case of transport measurements (first $\omega\rightarrow0$, then $\gamma\rightarrow0$), the universal conductivity $\sigma_{dc}=\frac{4}{\pi}\frac{e^2}{h}$ is thus observed,\cite{Miao07} whereas in optical experiments (first $\gamma\rightarrow0$, then $\omega\rightarrow0$) $\sigma_0=\frac{\pi}{2}\frac{e^2}{h}$ is seen.\cite{Mak08,Nair08} Clearly, it is the latter order of limits which is relevant in the context of plasmonics and we will always assume this universal value to be taken.

The local charge response of Eq. (\ref{LongWave}) contains intra- as well as interband contributions. To discuss longitudinal plasmons, mainly intraband transitions need to be considered. In the local approximation, the band-overlap goes to one, $f_+\to1$, and we can approximate for general isotropic dispersion
\begin{eqnarray}
\chi_{\rho\rho}&=&\frac{g_sg_v}{(2\pi)^2}\int d^2k\frac{n_F(E(\k))-n_F(E(\k+\q))}{\hbar\omega+E(\k)-E(\k+\q)}\;.
\end{eqnarray}
In the limit $q\rightarrow0$, this becomes
\begin{equation}
\chi_{\rho\rho}=\frac{g_sg_v}{(2\pi)^2}\int d^2k\left(-\frac{\partial n_F(E(\k))}{\partial E(\k)}\right)\left(\frac{\nabla E(\k)\cdot\q}{\hbar\omega}\right)^2,
\end{equation}
which can be evaluated for low temperatures, leading to Eq. (\ref{localCurrentResponse}).

\subsection{Current Response}
We now discuss the general current-current correlator. By this, we can treat longitudinal and transverse response on the same footing. We will thus use $\delta H=-q_ej^i A^i$ with $q_e=-e$ the electron charge ($e>0$) and $\psi^\alpha\to j^i$, $\lambda^\alpha\to q_eA^i$.\footnote{We use greek indices $\alpha,\beta$ for 3D and latin indices $i,j$ for 2D.}

The current operator for the full tight binding model needs to be defined with care because a simple Peierls substitution on the lattice breaks gauge invariance. Within an adequate continuum model, we obtain the general expression for the paramagnetic current-current correlation function\cite{Stauber10b} 
\begin{eqnarray}
\label{currentcorrelation}
\chi^{P;i,j}(\q,\omega)&=&\left(\frac{te}{\hbar}\right)^2\frac{g_s}{(2\pi)^2}\int_{{\rm{1.BZ}}}d^2k\sum_{s,s'=\pm}f_{s\cdot s'}^{i,j}(\k,\q)\nonumber\\
&\times&\frac{n_F(E^{s}({\bf k}))-n_F(E^{s'}({\bf k}+{\bf q}))}{E^{s}({\bf k})-E^{s'}({\bf k}+{\bf q})+\hbar\omega+\I0}\;,
\end{eqnarray}
with the same definitions as for the density-density correlation function below Eq. (\ref{densitycorrelation}), but the band-overlap is now given by
\begin{eqnarray}
\label{overlapp}
f_{\pm}^{i,j}(\k,\q)&=&\frac{1}{2}\Bigg(\re\left[\tilde\phi^i(\k,\q)(\tilde\phi^j(\k,\q))^*\right]\\\nonumber
&\pm&\re\left[\tilde\phi^i(\k,\q)\tilde\phi^j(\k,\q)\frac{\phi^*(\k)}{|\phi(\k)|}\frac{\phi^*(\k+\q)}{|\phi(\k+\q)|}\right]\Bigg)\;,\\
\tilde\phi^i(\k,\q)&=&\sum_\d\frac{\delta^i}{\q\cdot\d}\left(e^{\I(\k+\q)\cdot\d}-e^{\I\k\cdot\d}\right)\;.
\end{eqnarray}

The physical response, $\chi^{i,j}$,  also includes the diamagnetic contribution, 
\begin{eqnarray}
&&\chi^{i,j}(\q, \omega)=\chi^{P;i,j}(\q, \omega) + \chi^{D;i,j}_{\q}\;,\\
&&\chi^{D;i,j}_{\bm q}=\frac{e^2}{\hbar^2} \; h_{\text{bond}} \sum_{\d} \delta^i \delta^j 
\frac{4}{(\bm q \cdot \d)^2} \; \sin^2(\frac{\bm q \cdot \d}{2}) 
\;,\nonumber
\end{eqnarray}
where the energy per bond per unit area is given by
\begin{align}
\nonumber
h_{\text{bond}} =\frac{g_s}{3(2\pi)^2}\int_{\rm{1.BZ}}d^2k E^+(\k)\sum_{s=\pm}s\cdot n_F(E^{-s}({\bf k}))\;.
\end{align} 
Charge conservation then implies
\begin{equation}
\label{WardIdentity}
q_i \; \chi^{i,j}(\q, \omega) \; q_j=e^2\omega^2 \chi_{\rho\rho}(\bm q, \omega)
\;.\end{equation} 
Notice that the anisotropy of the response for finite $\q$ requires the  full
tensorial structure of $\chi^{i,j}$. The conductivity tensor is defined by $\sigma^{i,j}=\I\frac{e^2\chi^{i,j}}{\omega+\I0}$.

The system linearized around the Dirac point is rotationally invariant. We can thus decompose the response tensor $\chi^{i,j}$ into a longitudinal ($\chi_{jj}^+$) and transverse ($\chi_{jj}^-$) scalar component,
\begin{equation}
\chi^{i,j}(\q,\omega)=\frac{q_iq_j}{|\q|^2}\chi_{jj}^{+}(|\q|,\omega)+\left(\delta_{i,j}-\frac{q_iq_j}{|\q|^2}\right)\chi_{jj}^{-}(|\q|,\omega)\;.
\end{equation}
The longitudinal component is thus directly related to the polarizability via Eq. (\ref{WardIdentity}), i.e., $\chi_{jj}^+=\chi_{\rho\rho}\frac{\omega^2}{q^2}$. The transverse component, $\chi_{jj}^-$, was first calculated in Ref. \onlinecite{Principi09}.

Following Ref. \onlinecite{Wunsch06}, the results can be written in compact form using two dimensionless, complex functions defined as
\begin{eqnarray}
F^\pm(q,\omega)&=&\frac{g_sg_v}{16\pi}\frac{\hbar\omega}{t}\left[1-\left(\frac{v_Fq}{\omega}\right)^2\right]^{\mp\frac{1}{2}}\;,\\
G^\pm(x)&=&x\sqrt{x^2-1}\mp \ln\left(x+\sqrt{x^2-1}\right)\;,
\end{eqnarray}
with the Fermi velocity $\hbar v_F=\frac{3}{2}at$ and carbon-carbon distance $a=0.142$nm. We then can write
\begin{equation}
\label{CurrentCurrent}
\hbar^2\chi_{jj}^\pm(q,\omega)=t\left[\tilde\chi_0^\pm(q,\omega)+\tilde\chi_\mu^\pm(q,\omega)\right]\;,
\end{equation}
where the dimensionless functions $\tilde\chi_0^\pm=-\I\pi F^\pm(q,\omega )$ contain the response of the system at half-filling, i.e., genuine interband contributions. $\tilde\chi_\mu^\pm$ contains the additional contributions due to the finite chemical potential $\mu$,
\begin{eqnarray}
\label{CurrentCurrentMu}
&&\tilde\chi_\mu^\pm(q,\omega)=\mp\frac{g_sg_v}{2\pi}\frac{\mu}{t}\frac{\omega^2}{(v_Fq)^2}\pm F^\pm(q,\omega)\Big\{
G^\pm\left( x_+\right)\label{eq:DP}\\ 
&&-\Theta
\left( x_--1\right) \left[
G^\pm\left(x_-\right) \mp \I\pi
\right]-\Theta \left(1-x_-\right) G^\pm\left(-x_-\right) \Big\} \nonumber\;,
\end{eqnarray}
where we defined $x_\pm=\frac{2\mu\pm\hbar\omega}{\hbar v_{F}q}$. 

\begin{figure}
   \centering
   \includegraphics[height=0.5\columnwidth]{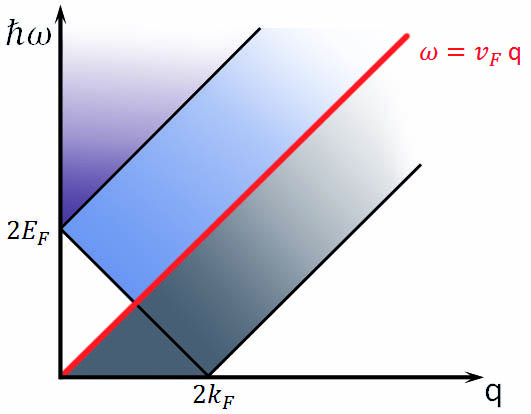} 
      \caption{Particle hole continuum of pristine graphene. Grey shaded regions correspond to intraband transitions, whereas blue and violet shaded regions indicate interband transitions. The white triangle with $\hbar\omega<2E_F$ indicates the area where long-lived plasmons may exist.}
   \label{fig:Diagram}
\end{figure}

Writing Eq. (\ref{CurrentCurrentMu}) in real and imaginary expressions, the results are divided in six  different zones, three above and three below the Dirac cone dispersion $\omega=v_Fq$, see Fig \ref{fig:Diagram}. The imaginary part is zero in the white areas and the left triangle is the Pauli-protected region where long-lived plasmons exist.

\subsection{Photon Propagator}
The retarded photon Green's function $\mathcal{D}^{\alpha\beta}$ for the gauge field is defined by $\psi^\alpha\to A^\alpha$ and $\lambda^\alpha\to q_ej^\alpha$. Within the Weyl gauge, i.e., setting the scalar potential to zero, $\phi=0$, it reads\cite{Stauber12}
\begin{equation}
\varepsilon_0\mathcal{D}^{\alpha\beta}({\bf k},\omega)=\frac{1}{\omega^ 2-c^ 2k^ 2}\left(\delta_{\alpha,\beta}-\frac{k^\alpha k^\beta}{\omega^2/c^2}\right)\;.
\end{equation} 
We will now derive the representations useful for two- and one-dimensional geometries. Part of this discussion can also be found in Ref. \onlinecite{Novotny06}. In the context of 2D graphene, the main equations were first derived in Ref. \onlinecite{Hanson08}.

\subsubsection{Two dimensional geometries}

In a layered structure, assumed to be perpendicular to the $z$ axis, the components parallel to the interface, $\q=(q_x,q_y)$, will be preserved as a good quantum number. It is, therefore, convenient to employ the following representation for the Green's function in a homogeneous medium
\begin{equation}\label{D0zz'}
 \mathcal{D}^{\alpha \beta}(z,z';\bm q,\omega) = \frac{1}{2\pi}
 \int dk_z \, {\rm e}^{\I k_z (z-z')} \, 
 \mathcal{D}^{\alpha \beta}(\bm k,\omega)
,\end{equation} 
with $\bm k = (\bm q,k_z)$.

In this representation, the tensor components have a different structure depending on whether $\alpha,\beta=i,j$ with $i,j=x,y$ or $\alpha,\beta=z$. For graphene plasmonics, we are mainly interested in the in-plane components. Decomposed into longitudinal and transverse contributions, we obtain with $q' = \sqrt{q^2-\epsilon\mu(\omega/c)^2}$
\begin{equation}\label{D0explicit}
 \mathcal{D}^{i j}= \left[ \frac{q_i q_j}{q^2}d^+(\bm q,\omega) +\left(\delta_{i j}-\frac{q_i q_j}{q^2}\right)d^-(\bm q,\omega) \right]{\rm e}^{-q'|z-z'|}\;,
\end{equation} 
where the in-plane longitudinal ($d^+$) and transverse ($d^-$) propagators are given by
\begin{equation}\label{dlt}
d^+ = \frac{q'}{2 \epsilon\varepsilon_0 \omega^2}\;, \;
d^- = -\frac{\mu\mu_0}{2q'}=-\frac{\mu}{2\varepsilon_0q'c^2}\;.
\end{equation} 
Above, we introduced the vacuum permittivity $\varepsilon_0$ and permeability $\mu_0$ as well as the relative dielectric constant $\epsilon$ and relative permeability $\mu$. Note that the negative transverse propagator can be obtained from the longitudinal one by the substitutions, $\epsilon\to q'$, $q'\to\mu$, and $\omega\to c$. This substitution is completed by $\chi_{jj}^+\to-\chi_{jj}^-$ and holds for all in-plane quantities. By considering in-plane components, it thus usually suffices to discuss the longitudinal channel.  

\subsubsection{One dimensional geometries}
To discuss the propagation in a nanowire, it is convenient to employ the following representation of the propagator:
\begin{equation}
 \mathcal{D}^{\alpha,\beta}({\bf r}-{\bf r'},q,\omega)=\frac{1}{(2\pi)^ 2}\int d^2p e^ {i\bf p\cdot({\bf r}-{\bf r'})}{D}^{\alpha,\beta}({\bf k},\omega)\;,
\end{equation} 
with ${\bf k}=({\bf p},q)$ and ${\bf R}=({\bf r},z)$. With the two-dimensional Green's function
\begin{equation}
 g({\bf r},q)=\frac{1}{(2\pi)^2}\int d^2p \frac{e^{i{\bf p}\cdot{\bf r}}}{p^ 2+q^ 2}=\frac{1}{2\pi}K_0(qr)\;,
\end{equation} 
where $K_0$ denotes the modified Bessel function of the third kind, the one dimensional photon propagator is defined by $g$ in the limit $r\to0$. Since $K_0(x)\to-\ln x$, this limit is not well-defined and a regularization procedure is needed. If we model the nanowire by a small cylinder with radius $a$ and look for the average of the fields induced by a uniform perturbation, we need to consider
\begin{equation}
\mathcal{D}^ {\alpha,\beta}(q,\omega)=\frac{1}{\pi a^2}\int d^2r\int d^2r' \mathcal{D}^{\alpha,\beta}({\bf r}-{\bf r'},q,\omega)\;.
\end{equation} 
This yields the following (well-defined) longitudinal ($d_1^+=\mathcal{D}^{z,z}$) and transverse ($d_1^-=\mathcal{D}^{x,x}=\mathcal{D}^{y,y}$) one-dimensional propagator
\begin{eqnarray}
\varepsilon_0d_1^+&=&\frac{1}{2\pi}\frac{q'^2}{\omega^2}\ln\frac{\sqrt{e}}{q'a}\;,\\\label{App:TransProp}
\varepsilon_0d_1^-&=&\frac{1}{2\pi a^ 2}\left[\frac{1}{\omega^2}-\left(\frac{a^ 2}{c^2}+\frac{(q'a)^2}{2\omega^2}\right)\ln\frac{\sqrt{e}}{q'a}\right]\;,
\end{eqnarray}
with the retarded wave number $q'=\sqrt{q^2-\omega^2/c^2}$. All other tensor components are zero.

The longitudinal propagator will give rise to plasmons which disperse as $q'\sqrt{-\ln(q'a)}$,\cite{Brey07} whereas the transverse propagator will predominantly give rise to transverse, but charged plasmons.\cite{Christensen12} But in the limit $q'\rightarrow0$, $d_1^-<0$ and there will be purely light-like transverse plasmons. These are discussed in more detail in Sec. \ref{Sec:TransversePlasmonsWire}.

\subsection{Graphene-Light coupling}
Within the Dirac cone approximation, the linear response of the gauge field and the current  decomposes into a longitudinal and transverse channel. We can thus write the response as
\begin{equation}
\delta A^\pm=e\chi_{jj}^\pm\delta j^\pm\;,\;\delta j^\pm=ed^\pm\delta A^\pm\;,
\end{equation}
with the photon propagator $d^\pm$ and the current propagator $\chi_{jj}^\pm$, respectively. Inserting one equation into the other, we obtain self-sustained oscillations for
\begin{equation}
1-e^2d^\pm\chi_{jj}^\pm=0\;.
\end{equation}
Within the Dirac cone approximation, longitudinal and transverse plasmonic excitations are thus decoupled.
 
The product of $e^2d^\pm$ and $\chi_{jj}^\pm$ is dimensionless. The bosonic response function $e^2d^\pm$ should therefore be related to the inverse of the fermonic response function $\chi_{jj}^{\pm}$. In the case of zero chemical potential and for the photonic propagator in vacuum, this relation becomes particularly clear. From the previous equations, we have
\begin{eqnarray}
\chi_{jj}^\pm &=& \mp\frac{g_s g_v\omega}{16 \hbar} \left[\left(\frac{v_Fq}{\omega}\right)^2-1\right]^{\mp1/2}\;,\\
e^2d^\pm&=& \pm\frac{2\pi\alpha\hbar}{\omega}\left[\left(\frac{cq}{\omega}\right)^2-1\right]^{\pm1/2}\;,
\end{eqnarray} 
with $\alpha$ the fine-structure constant. We thus obtain a dualism between massless bosons and massless (Dirac) fermions, $e^2d^\pm\leftrightarrow1/\chi_{jj}^\pm$, by interchanging $c\leftrightarrow v_F$ and $2\pi\alpha\leftrightarrow-16/(g_sg_v)$. For a homogeneous dielectric medium with general $\epsilon$ and $\mu$, the mapping is slightly different for longitudinal or transverse channel.

We finally comment on the case of propagating light $\omega<cq$ since taking the right branch cut requires some care. For $\omega\rightarrow\omega+i\delta$, we must have Im$d<0$ following our convention. This yields
\begin{equation}\label{dltPropagating}
e^2d^\pm = -\I\frac{2\pi\alpha\hbar}{\omega}\left[1-\left(\frac{cq}{\omega}\right)^2\right]^{\pm1/2}\;.
\end{equation} 

\bibliography{plasmonics} % Produces the bibliography via BibTeX.
\end{document}